\documentclass[vecphys]{svmult}

\usepackage{makeidx}         % allows index generation
\usepackage{graphicx}        % standard LaTeX graphics tool
\usepackage{multicol}        % used for the two-column index
\usepackage[bottom]{footmisc}% places footnotes at page bottom
\makeindex             % used for the subject index
\begin{document}
\newcommand{\newc}{\newcommand}

\newc{\be}{\begin{equation}}
\newc{\ee}{\end{equation}}
\newc{\ba}{\begin{eqnarray}}
\newc{\ea}{\end{eqnarray}}
\newc{\bea}{\begin{eqnarray*}}
\newc{\eea}{\end{eqnarray*}}
%\newc{\D}{\partial}
\newc{\ie}{{\it i.e.} }
\newc{\eg}{{\it e.g.} }
\newc{\etc}{{\it etc.} }
\newc{\etal}{{\it et al.}}
\newcommand{\nn}{\nonumber}

\newc{\ra}{\rightarrow}
\newc{\lra}{\leftrightarrow}
\newc{\lsim}{\buildrel{<}\over{\sim}}
\newc{\gsim}{\buildrel{>}\over{\sim}}

\def\apj{{Astroph.\@ J.\ }}
\def\mn{{Mon.\@ Not.\@ Roy.\@ Ast.\@ Soc.\ }}
\def\asta{{Astron.\@ Astrophys.\ }}
\def\aj{{Astron.\@ J.\ }}
\def\prl{{Phys.\@ Rev.\@ Lett.\ }}
\def\pd{{Phys.\@ Rev.\@ D\ }}
\def\nucp{{Nucl.\@ Phys.\ }}
\def\nat{{Nature\ }}
\def\plb {{Phys.\@ Lett.\@ B\ }}
\def \jetpl {JETP Lett.\ }

\title*{Accelerating Universe:Observational Status and Theoretical Implications}
\author{L. Perivolaropoulos}
\institute{Department of Physics, University of Ioannina, Greece \\
\texttt{e-mail: leandros@cc.uoi.gr}}
\maketitle

This is a pedagogical review of the recent observational data
obtained from type Ia supernova surveys that support the
accelerating expansion of the universe. The methods for the
analysis of the data are reviewed and the theoretical implications
obtained from their analysis are discussed.

\section{Introduction}
\label{intro} Recent distance-redshift surveys
\cite{perl98,riess98,tonry03,snobs,Riess:2004nr,Astier:2005qq} of
cosmologically distant Type Ia supernovae (SnIa) have indicated
that the universe has recently (at redshift $z\simeq 0.5$) entered
a phase of accelerating expansion. This expansion has been
attributed to a dark energy \cite{Sahni:2004ai} component with
negative pressure which can induce repulsive gravity and thus
cause accelerated expansion. The evidence for dark energy has been
indirectly verified by Cosmic Microwave Background (CMB)
\cite{Spergel03} and large scale structure \cite{Tegmark04}
observations.

The simplest and most obvious candidate for this dark energy is
the cosmological constant \cite{Sahni:1999gb} with equation of
state $w=\frac{p}{\rho}=-1$. The extremely fine tuned value of the
cosmological constant required to induce the observed accelerated
expansion has led to a variety of alternative models where the
dark energy component varies with time. Many of these models make
use of a homogeneous, time dependent minimally coupled scalar
field $\phi$ (quintessence\cite{Peebles:1987ek,Sahni:1999qe})
whose dynamics is determined by a specially designed potential
$V(\phi)$ inducing the appropriate time dependence of the field
equation of state $w(z) = {{p(\phi)} \over {\rho(\phi)}}$. Given
the observed $w(z)$, the quintessence potential can in principle
be determined. Other physically motivated models predicting late
accelerated expansion include modified %%
gravity\cite{Perrotta:1999am,Torres:2002pe,Nojiri:2003ni},
Chaplygin gas\cite{Kamenshchik:2001cp}, Cardassian
cosmology\cite{Freese:2002sq}, theories with compactified extra
dimensions\cite{Perivolaropoulos:2002pn,Perivolaropoulos:2003we},
braneworld models\cite{Sahni:2002dx} etc. Such cosmological models
predict specific forms of the Hubble parameter $H(z)$ as a
function of redshift $z$. The observational determination of the
recent expansion history $H(z)$ is therefore important for the
identification of the viable cosmological models.

The most direct and reliable method to observationally determine
the recent expansion history of the universe $H(z)$ is to measure
the redshift $z$ and the apparent luminosity of cosmological
distant indicators (standard candles) whose absolute luminosity is
known. The luminosity distance vs. redshift is thus obtained which
in turn leads to the Hubble expansion history $H(z)$.

The goal of this review is to present the methods used to
construct the recent expansion history $H(z)$ from SnIa data and
discuss the most recent observational results and their
theoretical implications. In the next section I review the method
used to determine $H(z)$ from cosmological distance indicators and
discuss SnIa as the most suitable cosmological standard candles.
In section 3 I show the most recent observational results for
$H(z)$ and discuss their possible interpretations other than
accelerating expansion. In section 4 I discuss some of the main
theoretical implications of the observed $H(z)$ with emphasis on
the various parametrizations of dark energy (the simplest being
the cosmological constant). The best fit parametrizations are
shown and their common features are pointed out. The physical
origin of models predicting the best fit form of $H(z)$ is
discussed in section 5 where I distinguish between minimally
coupled scalar fields (quintessence) and modified gravity
theories. An equation of state of dark energy with $w<-1$ is
obtained by a specific type of dark energy called {\it phantom
energy} \cite{Caldwell:1999ew}. This type of dark energy is faced
with theoretical challenges related to the stability of the
theories that predict it. Since however the SnIa data are
consistent with phantom energy it is interesting to investigate
the implications of such an energy. These implications are
reviewed in section 6 with emphasis to the Big Rip future
singularity implied by such models as the potential death of the
universe. Finally, in section 7 I review the future observational
and theoretical prospects related to the investigation of the
physical origin of dark energy and summarize the main conclusions
of this review.

\section{Expansion History from the Luminosity Distances of SnIa}
Consider a luminous cosmological object emitting at total power L
(absolute luminosity) in radiation within a particular wavelength
band. Consider also an observer (see Fig. 1) at a distance $d_L$
from the luminous object. In a static cosmological setup, the
power radiated by the luminous object is distributed in the
spherical surface with radius $d_L$ and therefore the intensity
$l$ (apparent luminosity) detected by the observer is \be
l=\frac{L}{4\pi  d_L^2} \ee
\begin{figure}
\centering
\includegraphics[bb=40 70 550 790,width=8.0cm,height=12.0cm,angle=-90]{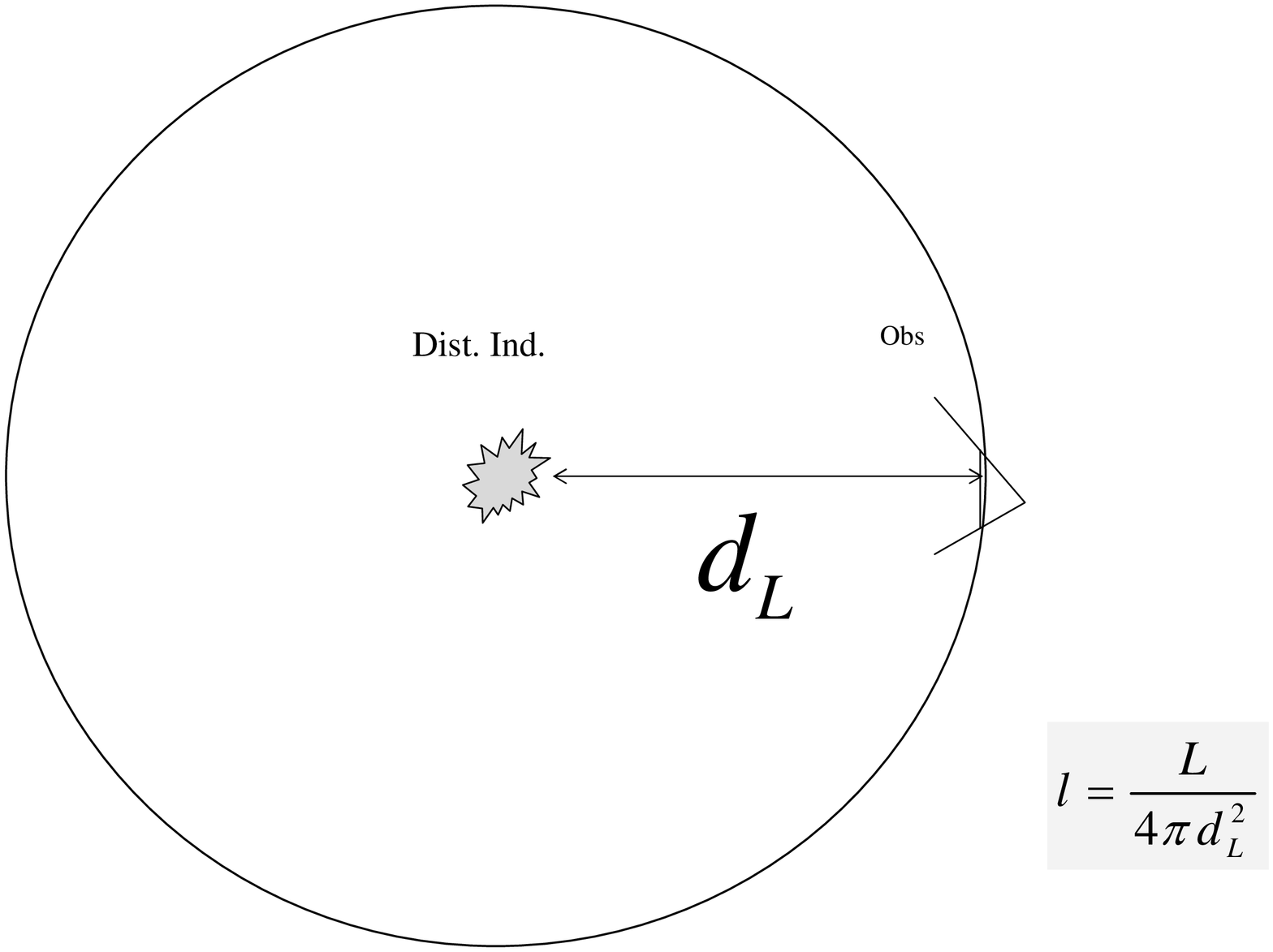}
%\picplace{5cm}{2cm} % Give the correct figure height and width in cm
%
\caption{The luminosity distance obtaned from the apparent and
absolute luminosities}
\label{fig:1}       % Give a unique label
\end{figure}

The quantity \be d_L \equiv \sqrt{\frac{L}{4\pi l}} \label{dll}\ee
is known as the luminosity distance to the luminous object and in
a static universe it coincides with the actual distance. In an
expanding universe however, the energy of the radiation detected
by the observer has been reduced not only because of the
distribution of photons on the spherical surface but also because
the energy of the photons has been redshifted while their
detection rate is reduced compared to their emission rate due to
the cosmological expansion \cite{kolbeu}. Both of these expansion
effects give a reduction of the detected energy by a factor
$\frac{a(t_0)}{a(t)}=(1+z)$ where $a(t)$ is the scale factor of
the universe at cosmic time $t$ and $t_0$ is the present time.
Usually $a$ is normalized so that $a(t_0)=1$. Thus the detected
apparent luminosity in an expanding background may be written as
\be l=\frac{L}{4\pi a(t_0)^2 x(z)^2 (1+z)^2} \ee where $x(z)$ is
the comoving distance to the luminus object emitting with redshift
$z$. This implies that in an expanding universe the luminosity
distance $d_L(z)$ is related to the comoving distance $x(z)$ by
the relation \be d_L(z)=x(z) (1+z) \label{dxrel} \ee Using eq.
(\ref{dxrel}) and the fact that light geodesics in a flat
expanding background obey \be c\; dt = a(z)\; dx(z)
\label{lgeod}\ee it is straightforward to eliminate $x(z)$ and
express the expansion rate of  the universe $H(z)\equiv \frac{\dot
a}{a}(z)$ at a redshift $z$ (scale factor $a=\frac{1}{1+z}$) in
terms of the observable luminosity distance as \be \label{hz1}
H(z)= c [{d\over {dz}} ({{d_L(z)}\over {1+z}})]^{-1}
\label{dlh}\ee This is an important relation that connects the
theoretically predictable Hubble expansion history $H(z)$ with the
observable luminosity distance $d_L(z)$ in the context of a
spatially flat universe. Therefore, if the absolute luminosity of
cosmologically distant objects is known and their apparent
luminosity is measured as a function of redshift, eq. (\ref{dll})
can be used to calculate their luminosity distance $d_L(z)$ as a
function of redshift. The expansion history $H(z)$ can then be
deduced by differentiation with respect to the redshift using eq.
(\ref{dlh}). Reversely, if a theoretically predicted $H(z)$ is
given, the corresponding predicted $d_L(z)$ is obtained from
(\ref{dlh}) by integrating $H(z)$ as \be d_L(z)=c\; (1+z) \int_0^z
\frac{dz'}{H(z')} \label{dlhr} \ee This predicted $d_L (z)$ can be
compared with the observed $d_L (z)$ to test the consistency of
the theoretical model with observations. In practice astronomers
do not refer to the ratio of absolute over apparent luminosity.
Instead they use the difference between apparent magnitude $m$ and
absolute magnitude $M$ which is connected to the above ratio by
the relation \be m-M=2.5\; log_{10}(\frac{L}{l}) \label{mmll} \ee

A particularly useful diagram which illustrates the expansion
history of the Universe is the {\it Hubble diagram}. The x-axis of
a Hubble diagram (see Fig. 2) shows the redshift $z$ of
cosmological luminous objects while the y-axis shows the physical
distance $\Delta r$ to these objects.
\begin{figure}
\centering
\includegraphics[bb=40 70 550 790,width=8.0cm,height=12.0cm,angle=-90]{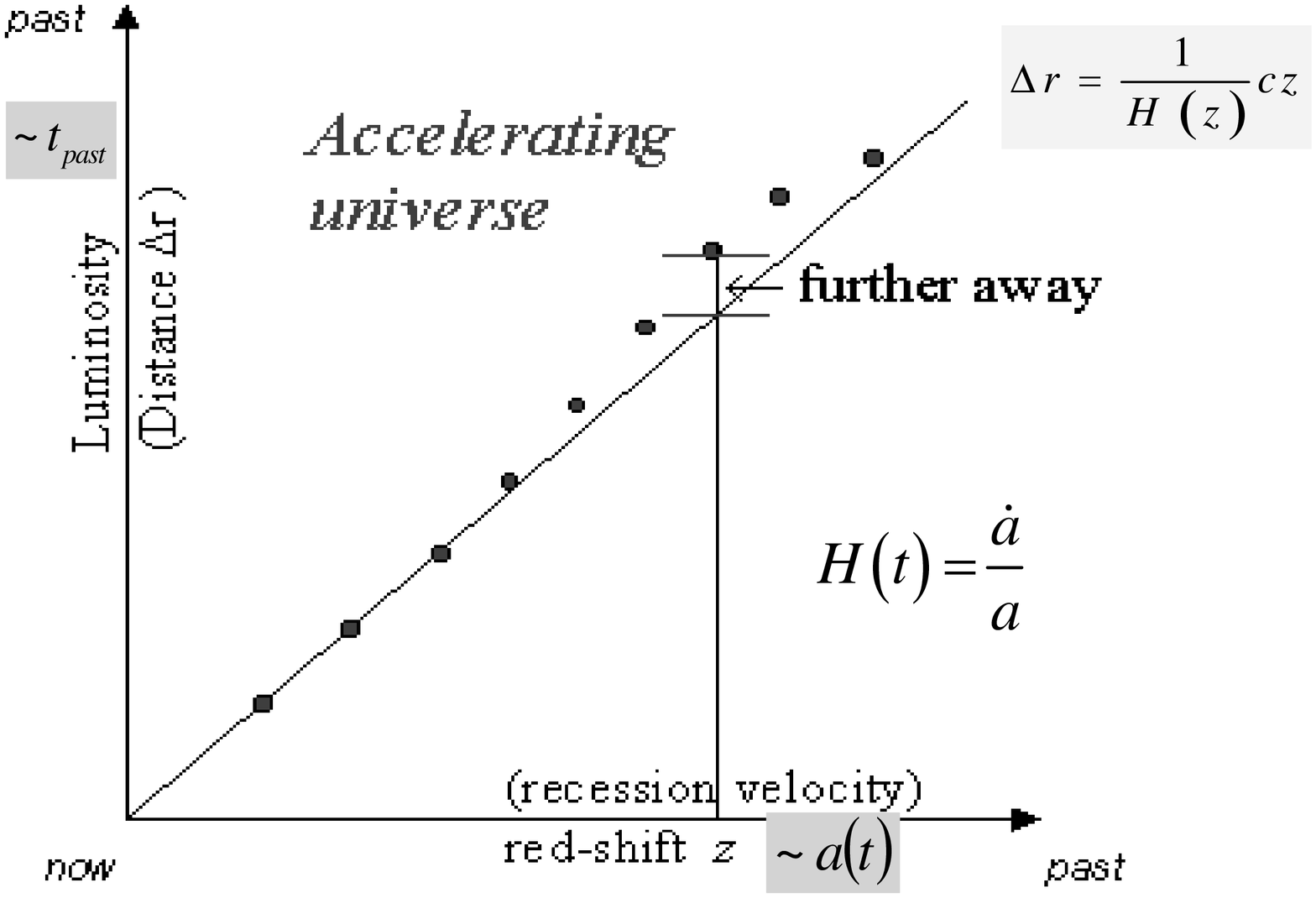}
%\picplace{5cm}{2cm} % Give the correct figure height and width in cm
%
\caption{The Hubble diagram. In an accelerating universe luminous
objects at a given redshift appear to be dimmer.}
\label{fig:2}       % Give a unique label
\end{figure}
In the context of a cosmological setup the redshift $z$ is
connected to the scale factor $a(t)$ at the time of emission of
radiation by $1+z=\frac{a(t_0)}{a(t)}$ where $t_0$ is the present
time. On the other hand, the distance to the luminous object is
related to the time in the past $t_{past}$ when the radiation
emission was made. Therefore, the Hubble diagram contains
information about the time dependence of the scale factor $a(t)$.
The slope of this diagram at a given redshift denotes the inverse
of the expansion rate $\frac{\dot a}{a}(z)\equiv H(z)$ ie \be
\Delta r = \frac{1}{H(z)} c\; z \ee In an accelerating universe
the expansion rate $H(z)$ was smaller in the past (high redshift)
and therefore the slope $H^{-1}$ of the Hubble diagram is larger
at high redshift. Thus, at given redshift, luminous objects appear
to be further away (dimmer)  compared to an empty universe
expanding with a constant rate (see Fig. 2).

The luminous objects used in the construction of the Hubble
diagram are objects whose absolute luminosity is known and
therefore their distance can be evaluated from their apparent
luminosity along the lines discussed above. Such objects are known
as {\it distance indicators} or {\it standard candles}. A list of
common distance indicators used in astrophysics and cosmology is
shown in Table 1 along with the range of distances where these
objects are visible and the corresponding accuracy in the
determination of their absolute magnitude.
\begin{table}
\centering \caption{Extragalactic distance indicators (from Ref.
\cite{hogan97})}
\label{tab:1}       % Give a unique label
\begin{tabular}{lll}
\hline\noalign{\smallskip}
Technique & Range of distance & Accuracy ($1\sigma$)  \\
\noalign{\smallskip}\hline\noalign{\smallskip}
Cepheids & $<$ LMC to 25 Mpc &0.15 mag \\
SNIa & 4 Mpc to $>$ 2 Gpc & 0.2 mag \\
Expand. Phot. Meth./SnII& LMC to 200 Mpc& 0.4 mag \\
Planetary Nebulae & LMC to 20 Mpc& 0.1 mag \\
Surf. Brightness Fluct & 1 Mpc to 100 Mpc & 0.1 mag \\
Tully Fisher & 1 Mpc to 100 Mpc & 0.3 mag \\
Brightest Cluster Gal. & 50 Mpc to 1 Gpc & 0.3 mag \\
Glob. Cluster Lum. Fun. & 1 Mpc to 100 Mpc & 0.4 mag \\
Sunyaev-Zeldovich & 100 Mpc to $>$ 1 Gpc & 0.4 mag \\
Gravitational Lensing& ~5 Gpc & 0.4 mag \\
\noalign{\smallskip}\hline
\end{tabular}
\end{table}
As shown in Table 1 the best choice distance indicators for
cosmology are SnIa not only because they are extremely luminous
(at their peak they are as luminous as a bright galaxy) but also
because their absolute magnitude can be determined at a high
accuracy.

Type Ia supernovae emerge in binary star systems where one of the
companion stars has a mass below the Chandrasekhar limit $1.4
M_\odot$ and therefore ends up (after hydrogen and helium burning)
as white dwarf supported by degeneracy pressure. Once the other
companion reaches its red giant phase the white dwarf begins
gravitational striping of the outer envelop of the red giant thus
accreting matter from the companion star. Once the white dwarf
reaches a mass equal to the Chandrasekhar limit, the degeneracy
pressure is unable to support the gravitational pressure, the
white dwarf shrinks and increases its temperature igniting carbon
fussion. This leads to violent explosion which is detected by a
light curve which rapidly increases luminosity in a time scale of
less than a month, reaches a maximum and disappears in a timescale
of 1-2 months (see Fig. 3).
\begin{figure}
\centering
\includegraphics[bb=40 70 550 790,width=8.0cm,height=12.0cm,angle=-90]{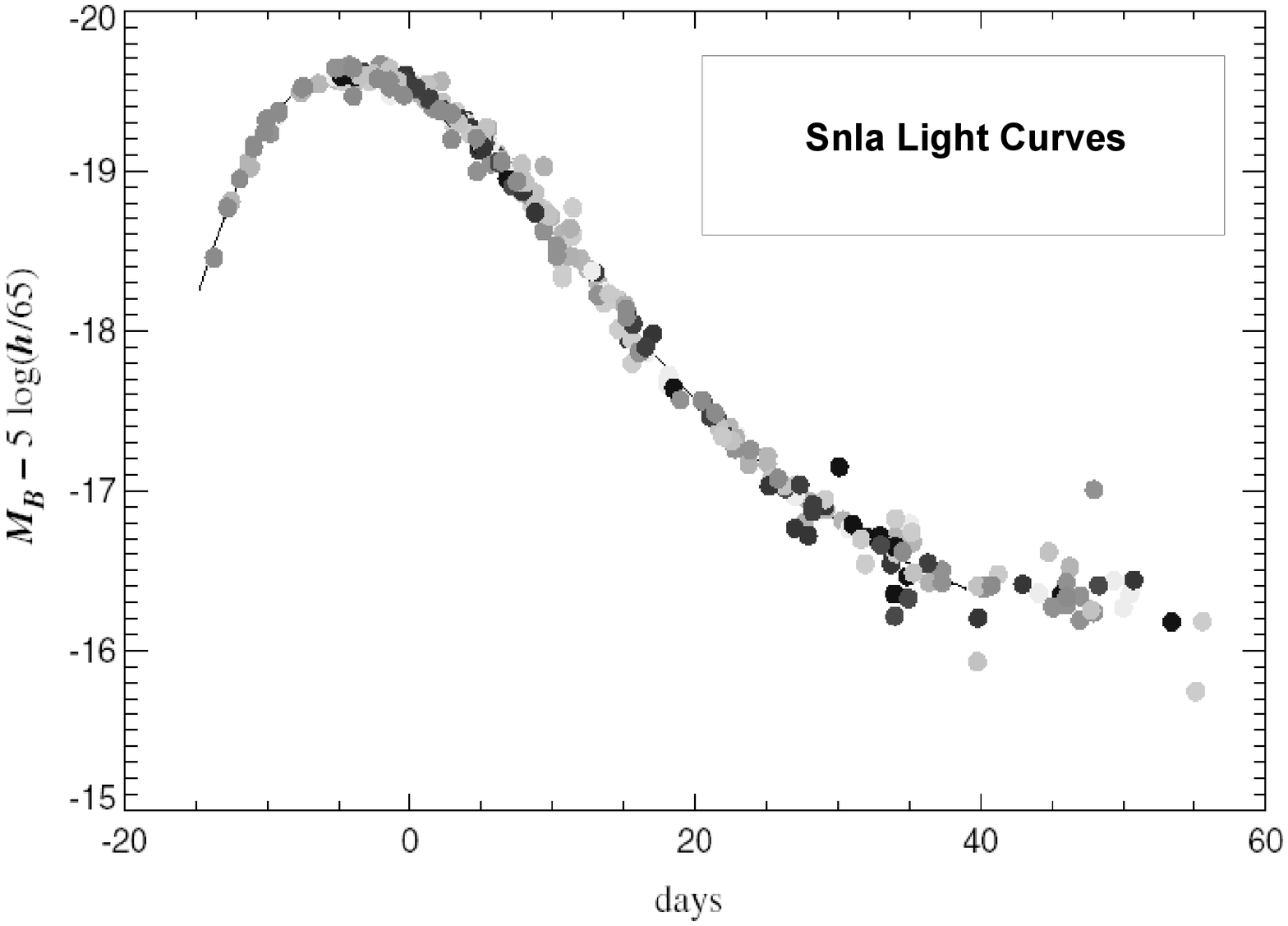}
%\picplace{5cm}{2cm} % Give the correct figure height and width in cm
%
\caption{Typical SnIa light-curve.}
\label{fig:3}       % Give a unique label
\end{figure}
Type Ia are the preferred distance indicators for cosmology for
several reasons:
\begin{enumerate}
\item They are exceedingly luminous. At their peak luminosity they
reach an absolute magnitude of $M\simeq -19$ which corresponds to
about $10^{10} M_\odot$. \item They have a relatively small
dispersion of peak absolute magnitude. \item Their explosion
mechanism is fairly uniform and well understood. \item There is no
cosmic evolution of their explosion mechanism according to known
physics. \item There are several local SnIa to be used for testing
SnIa physics and for calibrating the absolute magnitude of distant
SnIa.
\end{enumerate}
On the other hand, the main problem for using SnIa as standard
candles is that they are not easy to detect and it is impossible
to predict a SnIa explosion. In fact the expected number of SnIa
exploding per galaxy is 1-2 per millenium. It is therefore
important to develop a {\it search strategy} in order to
efficiently search for SnIa at an early stage of their light
curve. The method used (with minor variations) to discover and
follow up photometrically and spectroscopically SnIa consists of
the following steps \cite{perl98,riess98,tonry03,snobs}:
\begin{enumerate}
\item Observe a number of wide fields of apparently empty sky out
of the plane of our Galaxy. Tens of thousands of galaxies are
observed in a few patches of sky. \item Come back three weeks
later (next new moon) to observe the same galaxies over again.
\item Subtract images to identify on average 12-14 SnIa. \item
Schedule in advance follow up photometry and spectroscopy on these
SnIa as they brighten to peak and fade away.
\end{enumerate}
Given the relatively short time difference (three weeks) between
first and second observation, most SnIa do not have time to reach
peak brightness so almost all the discoveries are pre-maximum.
This strategy turns a rare, random event into something that can
be studied in a systematic way. This strategy is illustrated in
Figs 4 and 5 (from Ref. \cite{perlmuter97}).
\begin{figure}
\centering
\includegraphics[bb=40 70 550 790,width=8.0cm,height=12.0cm,angle=-90]{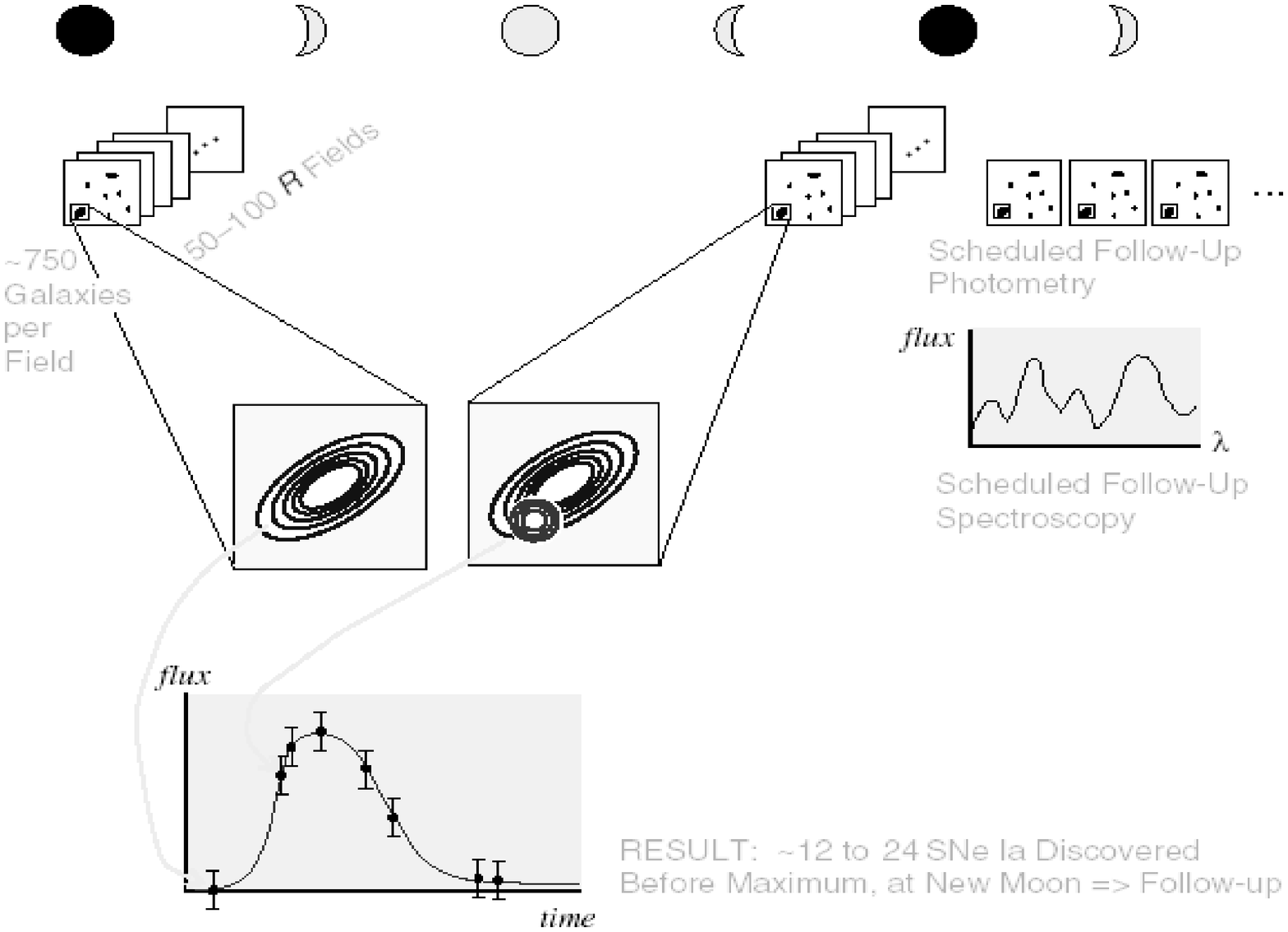}
%\picplace{5cm}{2cm} % Give the correct figure height and width in cm
%
\caption{Search strategy to discover of supernovae in a scheduled,
systematic procedure \cite{perlmuter97}}
\label{fig:4}       % Give a unique label
\end{figure}
The outcome of this observation strategy is a set of SnIa light
curves in various bands of the spectrum (see Fig. 6). These light
curves are very similar to each other and their peak apparent
luminosity could be used to construct the Hubble diagram assuming
a common absolute luminosity.

\begin{figure}
\centering
\includegraphics[bb=40 70 550 790,width=8.0cm,height=12.0cm,angle=-90]{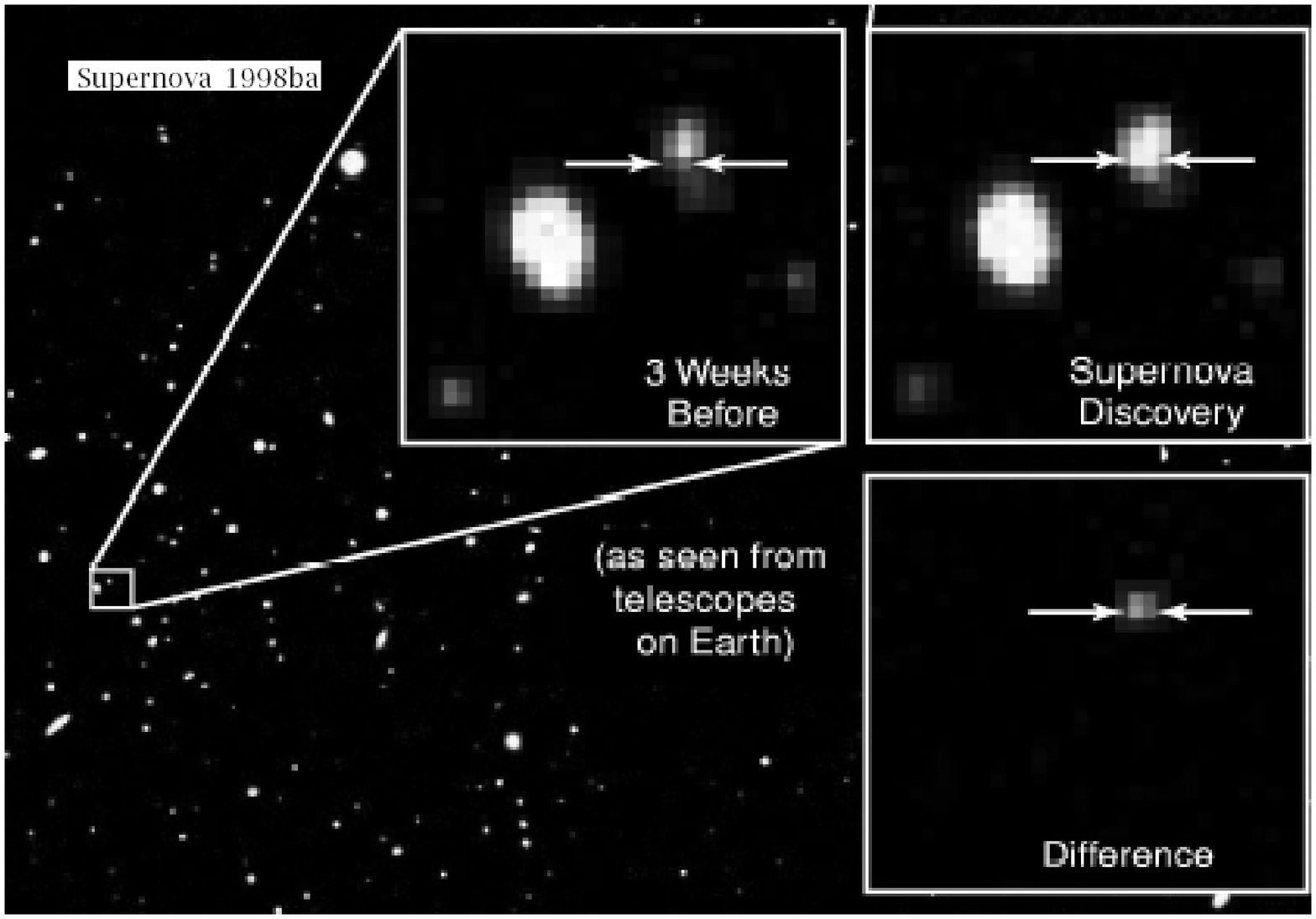}
%\picplace{5cm}{2cm} % Give the correct figure height and width in cm
%
\caption{Supernova 1998ba, an example of a supernova discovery
using the search strategy described in the text involving
subtraction of images.}
\label{fig:5}       % Give a unique label
\end{figure}

Before this is done however a few corrections must be made to take
into account the minor intrinsic absolute luminosity differences
(due to composition differences) among SnIa as well as the
radiation extinction due to the intergalactic medium. Using
samples of closeby SnIa it has been empirically observed that the
minor differences of SnIa absolute luminosity are connected with
differences in the shape of their light curves. Broad slowly
declining light curves (stretch factor $s>1$) correspond to
brighter SnIa while narrower rapidly declining light curves
(stretch factor $s<1$) correspond to intrinsically fainter SnIa.
\begin{figure}
\centering
\includegraphics[bb=40 70 550 790,width=8.0cm,height=12.0cm,angle=-90]{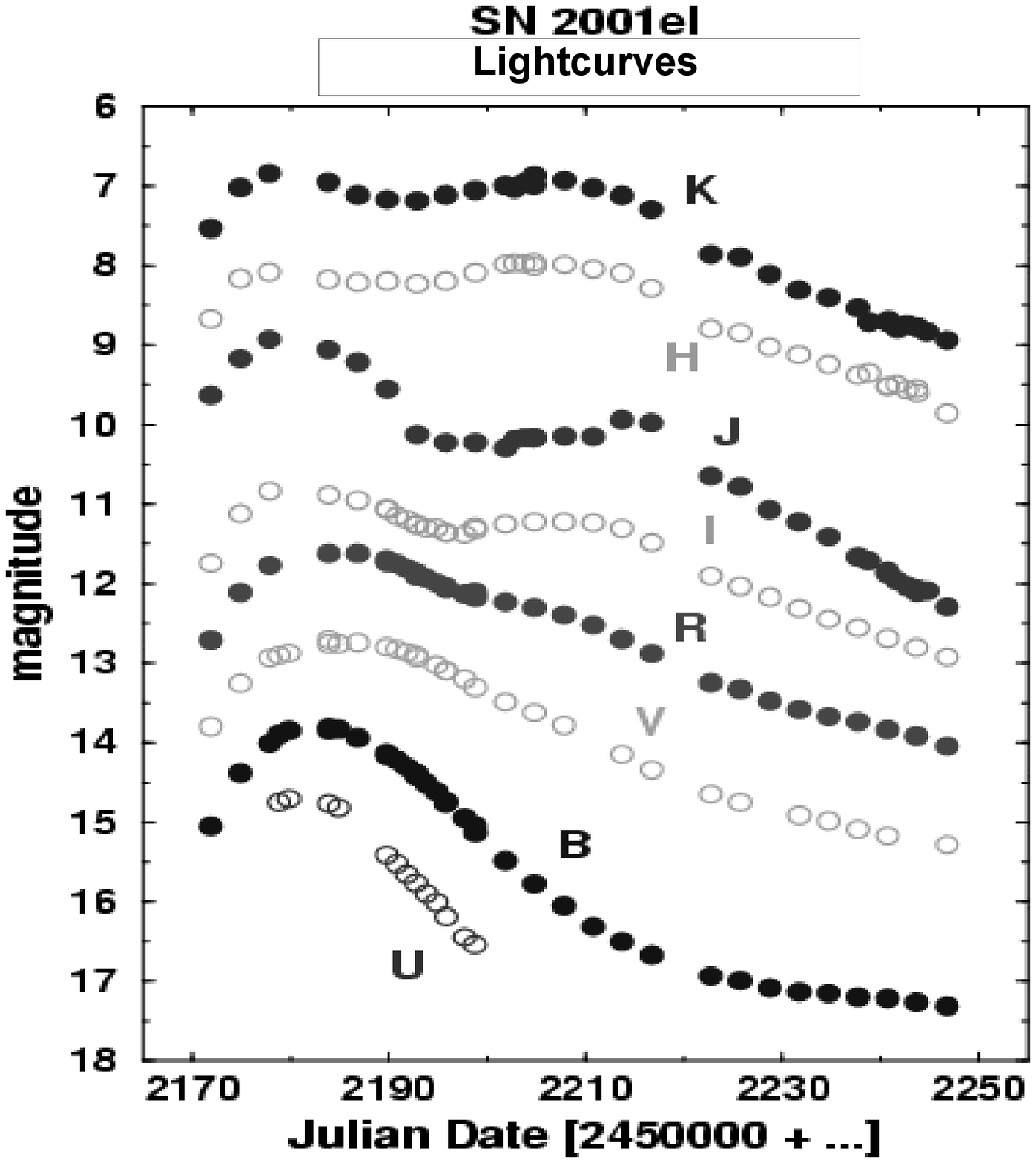}
%\picplace{5cm}{2cm} % Give the correct figure height and width in cm
%
\caption{A set of light curves from SN2001el in various bands of
the spectrum.}
\label{fig:6}       % Give a unique label
\end{figure}
This stretch factor dependence of the SnIa absolute luminosity has
been verified using closeby SnIa \cite{perl97} It was shown that
contraction of broad light curves while reducing peak luminosity
and stretching narrow light curves while increasing peak
luminosity makes these light curves coincide (see Fig. 7).
\begin{figure}
\centering
\includegraphics[bb=40 100 420 790,width=8.0cm,height=12.0cm,angle=-90]{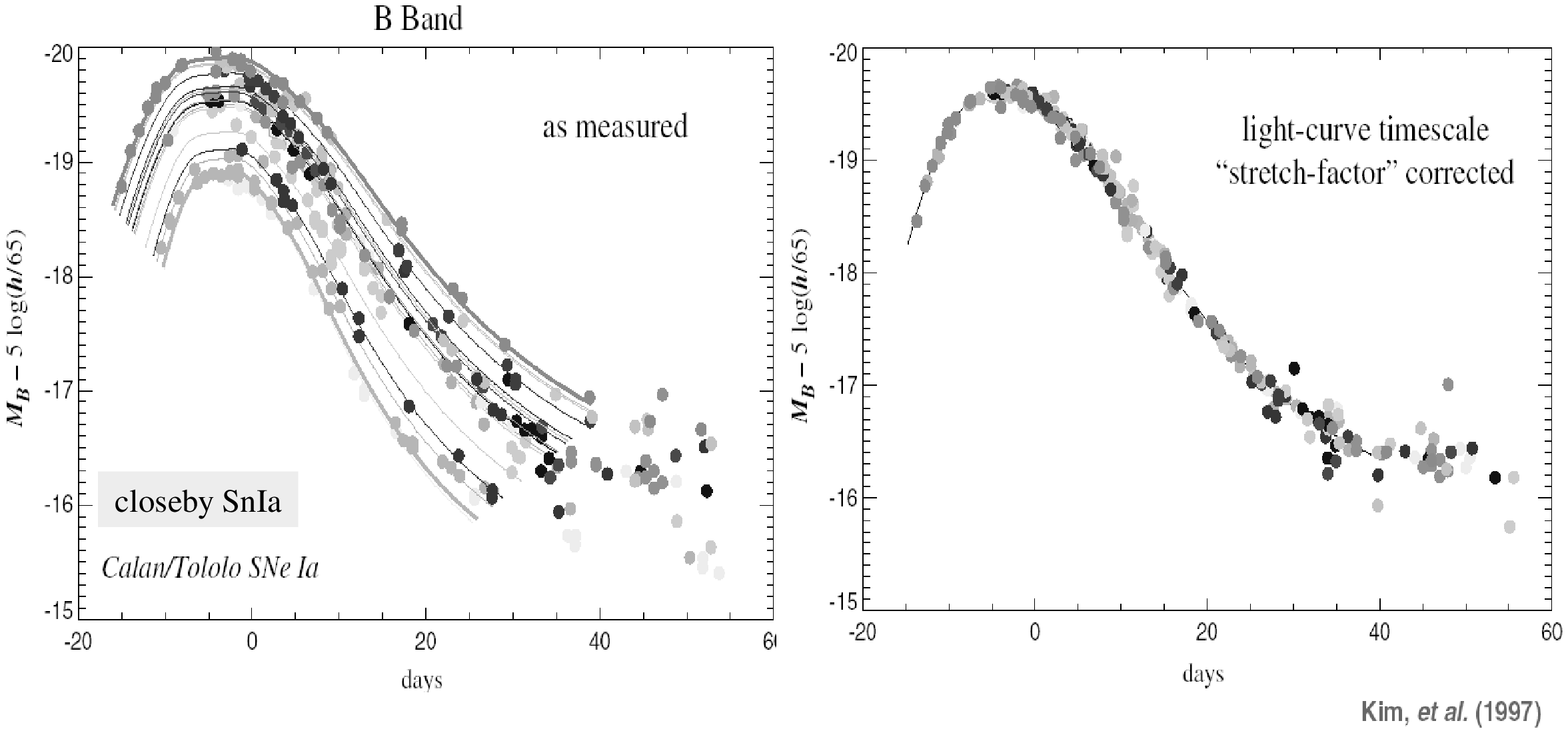}
%\picplace{5cm}{2cm} % Give the correct figure height and width in cm
%
\caption{Left:The range of lightcurve for low-redshift supernovae
discovered by the Calan/Tololo Supernova Survey. At these
redshifts, the relative distances can be determined (from
redshift), so their relative brightnesses are known. Right: The
same lightcurves after calibrating the supernova brightness using
the "stretch" of the timescale of the lightcurve as an indicator
of brightness (and the color at peak as an indicator of dust
absorption)}
\label{fig:7}       % Give a unique label
\end{figure}
In addition to the stretch factor correction an additional
correction must be made in order to compare the light curves of
high redshift SnIa with those of lower redshift. In particular all
light curves must be transformed to the same reference frame and
in particular the rest frame of the SnIa. For example a low
redshift light curve of the blue B band of the spectrum should be
compared with the {\it appropriate} red R band light curve of a
high redshift SnIa. The transformation also includes correction
for the cosmic time dilation (events at redshift $z$ last $1+z$
times longer than events at $z\simeq 0$).
\begin{figure}
\centering
\includegraphics[bb=40 70 500 790,width=8.0cm,height=12.0cm,angle=-90]{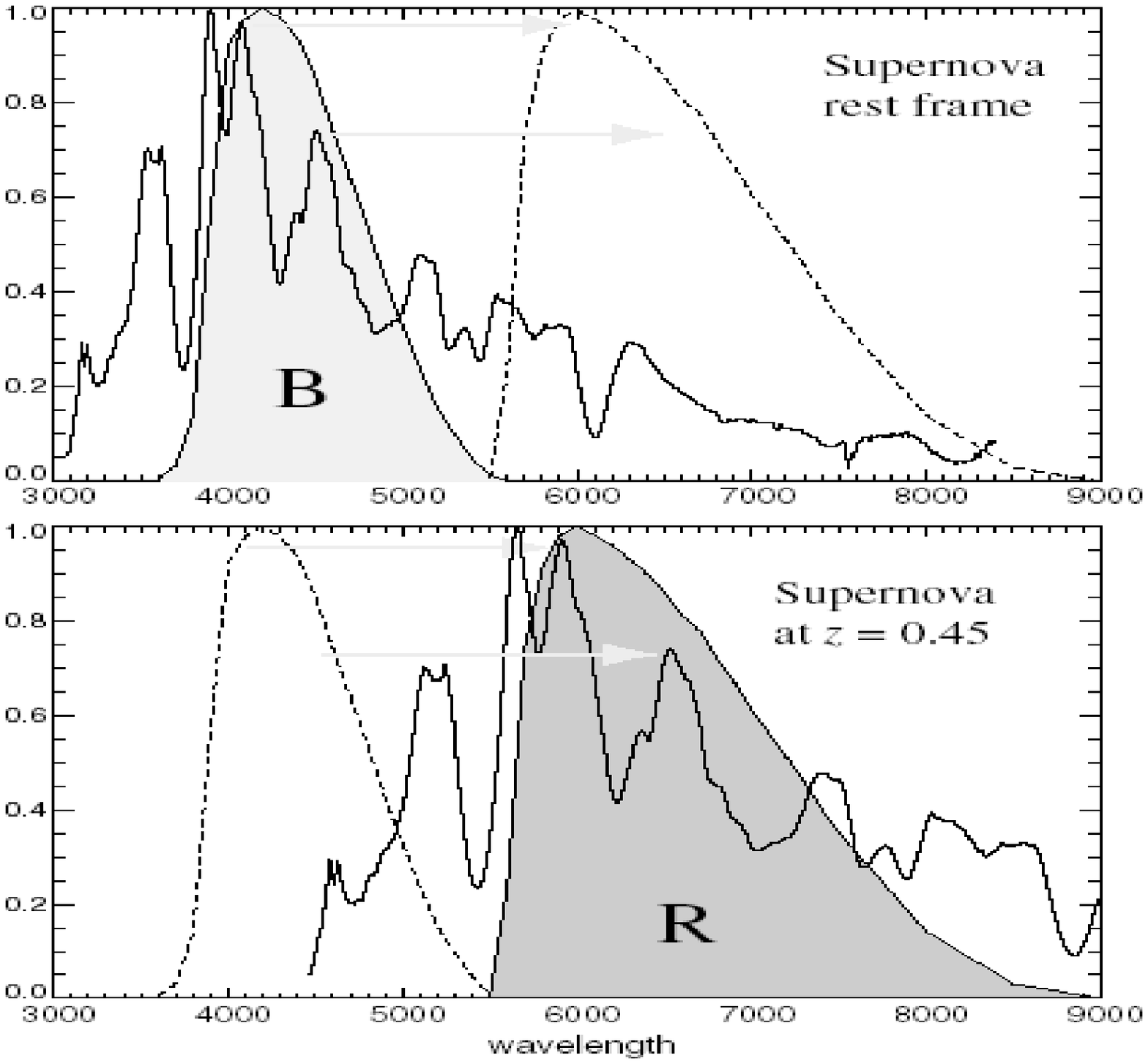}
%\picplace{5cm}{2cm} % Give the correct figure height and width in cm
%
\caption{Slightly different parts of the supernova spectrum are
observed through the "B filter" transmission function at low
redshift (upper panel) and through the "R filter" transmission
function at high redshift (lower panel). This small difference is
accounted for by the `cross-filter K-correction'\cite{Kim}.}
\label{fig:8}       % Give a unique label
\end{figure}
These corrections consist the {\it K-correction} and is used in
addition to the stretch factor correction discussed above. The
K-correction transformation is illustrated in Fig.8.

\section{Observational Results}
The first project in which SnIa were used to determine the
cosmological constant energy was the research from Perlmutter et
al. in 1997 \cite{perl97}. The project was known as the Supernova
Cosmology Project (SCP). Applying the above described methods they
discovered seven distant SnIa at redshift $0.35 < z < 0.65$. When
discovered, the supernovae were followed for a year by different
telescopes on earth to obtain good photometry data in different
bands, in order to measure good magnitudes. The Hubble diagram
they constructed was consistent with standard Friedman cosmology
without dark energy or cosmological constant.

A year after their first publication, Perlmutter et al. published
in Nature \cite{perl98} an update on their initial results. They
had included the measurements of a very high-redshifted z = 0.83
Supernova Ia. This dramatically changed their conlusions. The
standard decelerating Friedman cosmology was rulled out at about
$99\%$ confidence level. The newly discovered Supernova indicated
a universe with accelerating expansion dominated by dark energy.
These results were confirmed independently by another pioneer
group (High-z Supernova Search Team (HSST)) searching for SnIa and
measuring the expansion history $H(z)$  (Riess et al. in 1998
\cite{riess98}). They had discovered 16 SnIa at $0.16 < z < 0.62$
and their $H(z)$ also indicated accelerating expansion ruling out
for a flat universe. Their data also permitted them to definitely
rule out decelerating Friedman cosmology at about $99\%$
confidence level.

In 2003 Tonry et al. \cite{tonry03} reported the results of their
observations of eight newly discovered SnIa. These SnIa were found
in the region $0.3 < z < 1.2$. Together with previously acquired
SnIa data they had a data set of more than 100 SnIa. This dataset
confirmed the previous findings of accelerated expansion and gave
the first hints of decelerated expansion at redshifts $z\gsim 0.6$
when matter is expected to begin dominating over dark energy. This
transition from decelerating to accelerating expansion was
confirmed and pinpointed accurately by Riess et al. in 2004
\cite{Riess:2004nr} who included in the analysis 16 new
high-redshift SnIa obtained with HST and reanalyzed all the
available data in a uniform and robust manner constructing a
robust and reliable dataset consisting of 157 points known as the
Gold dataset. These SnIa included 6 of the 7 highest redshift SnIa
known with $z > 1.25$. With these new observations, they could
clearly identify the transition from a decelerating towards an
accelerating universe to be at $z = 0.46 \pm 0.13$. It was also
possible to rule out the effect of dust on the dimming of distant
SnIa, since the accelerating/decelerating transition makes the
effect of dimming inverse.
\begin{figure}
\centering
\includegraphics[bb=40 70 550 790,width=8.0cm,height=12.0cm,angle=-90]{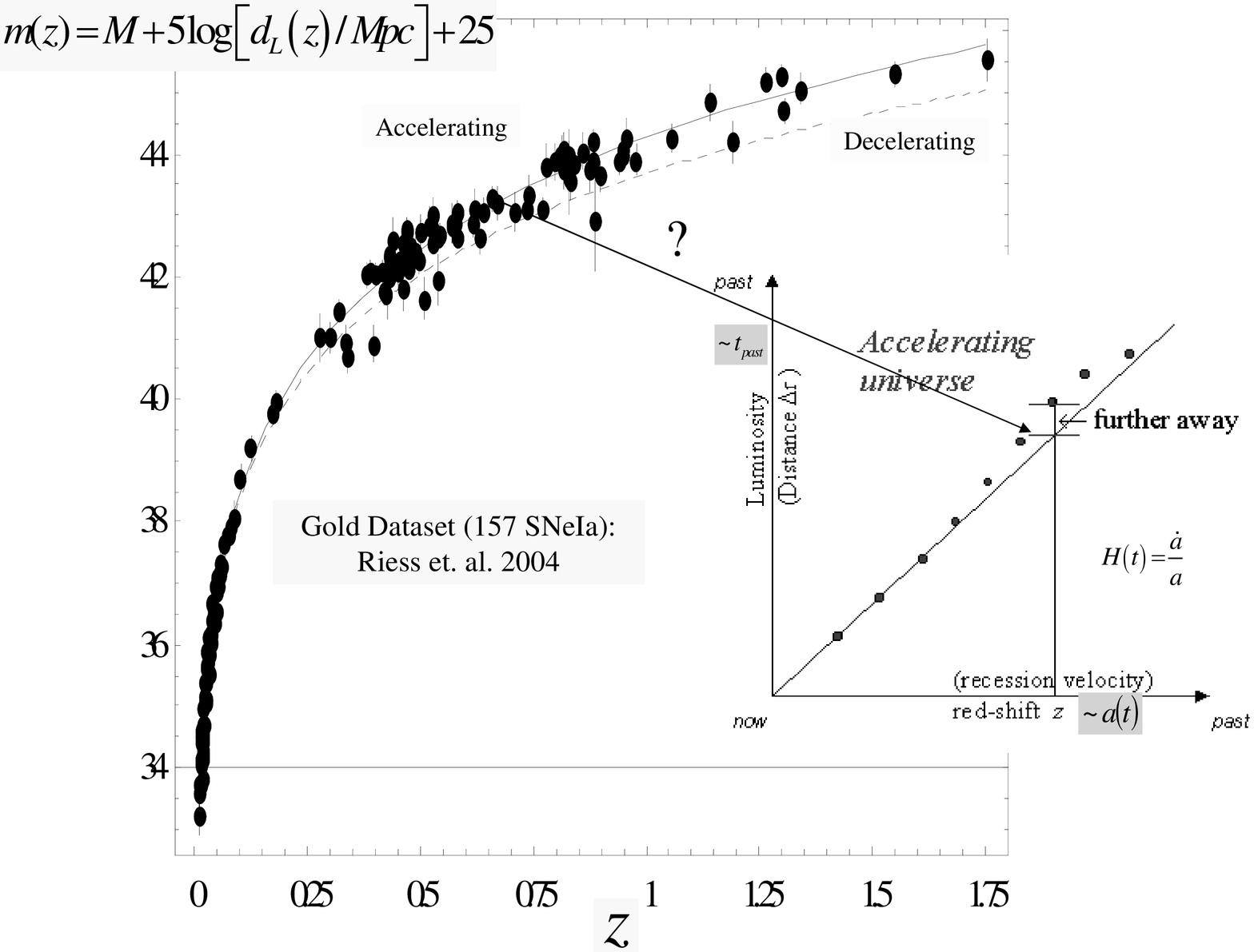}
%\picplace{5cm}{2cm} % Give the correct figure height and width in cm
%
\caption{The apparent magnitude $m(z)$ vs redshift as obtained
from the Gold dataset. It is not easy to distinguish between
accelerating and decelerating expansion in such a diagram.}
\label{fig:9}       % Give a unique label
\end{figure}
The Hubble diagram obtained from the Gold dataset is shown in Fig.
9 where the corrected apparent magnitude $m(z)$ of the 157 SnIa is
plotted versus the redshift $z$. The apparent magnitude $m(z)$ is
related to the corresponding luminosity distance $d_L$ of the SnIa
by \be m(z)=M + 5 log_{10}[ \frac{d_L(z)}{Mpc}]+25 \ee where $M$
is the absolute magnitude which is assumed to be constant for
standard candles like SnIa after the corrections discussed in
section 2 are implemented.

A potential problem of plots like the one of Fig. 9 is that it is
not easy to tell immediately if the data favor an accelerating or
decelerating universe. This would be easy to tell in the Hubble
diagram of Fig. 2 where the distance is plotted vs redshift and is
superposed with the distance-redshift relation $d_L^{empty}(z)$ of
an empty universe with $H(z)$ constant. An even more efficient
plot for such a purpose would be the plot of the ratio
$\frac{d_L(z)}{d_L^{empty}(z)}$ (or its $log_{10}$) which can
immediately distinguish accelerating from decelerating expansion
by comparing with the $\frac{d_L(z)}{d_L^{empty}(z)}=1$ line. Such
a plot is shown in Fig. 10 \cite{Riess:2004nr} using both the raw
Gold sample data and the same data binned in redshift bins.
\begin{figure}
\centering
\includegraphics[bb=40 70 550 790,width=8.0cm,height=12.0cm,angle=-90]{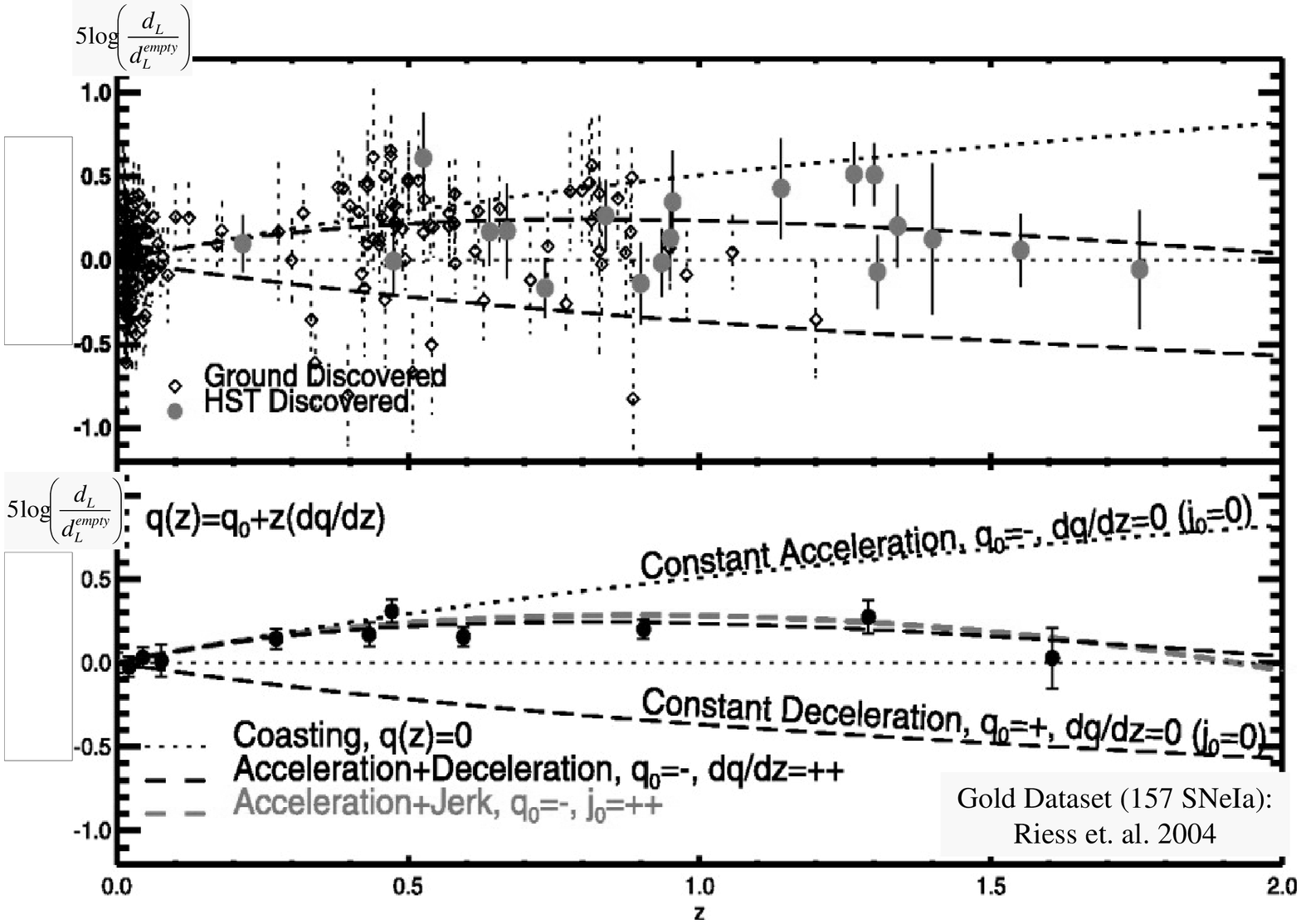}
%\picplace{5cm}{2cm} % Give the correct figure height and width in cm
%
\caption{The reduced Hubble diagram used to distinguish between
accelerating and decelerating expansion\cite{Riess:2004nr}.}
\label{fig:10}       % Give a unique label
\end{figure}

The lines of zero acceleration, constant acceleration and constant
deceleration are also shown for comparison. Clearly the best fit
is obtained by an expansion which is accelerating at recent times
($z\lsim 0.5$) and decelerating at earlier times ($z\gsim 0.5$)
when matter is expected to dominate.

The interpretation of the data assuming that the observed dimming
at high redshift is due to larger distance may not be the only
possible interpretation. The most natural alternative
interpretations however have been shown to lead to inconsistencies
and none of them has been favored as a viable alternative at
present. These alternative interpretations include the following:
\begin{itemize} \item{\bf Intergalactic Dust:}
Ordinary astrophysical dust does not obscure equally at all
wavelengths, but scatters blue light preferentially, leading to
the well-known phenomenon of ``reddening''. Spectral measurements
\cite{Riess:2004nr} reveal a negligible amount of reddening,
implying that any hypothetical dust must be a novel ``grey''
variety inducing no spectral distortions \cite{Aguirre:1998ge}.
\item {\bf Grey Dust:} Grey dust is highly constrained by
observations: first, it predicts further increase of dimming at
higher redshifts $z\gsim 0.5$ which is not observed; and second,
intergalactic dust would absorb ultraviolet/optical radiation and
re-emit it at far infrared wavelengths, leading to stringent
constraints from observations of the cosmological far-infrared
background. Thus, while the possibility of obscuration has not
been entirely eliminated, it requires a novel kind of dust which
is already highly constrained (and may be convincingly ruled out
by further observations). \item {\bf Evolution of SnIa:} The
supernova search teams have found consistency in the spectral and
photometric properties of SnIa over a variety of redshifts and
environments \cite{Riess:2004nr} (e.g., in elliptical vs. spiral
galaxies). Thus despite the relevant tests there is curently no
evidence that the observed dimming can be attributed to evolution
of SnIa.
\end{itemize}

According to the best of our current understanding, the supernova
results indicating an accelerating universe seem likely to be
trustworthy. Needless to say, however, the possibility of a
 neglected systematic effect can not be definitively excluded.
 Future experiments, discussed in section 7
will both help us improve our understanding of the physics of
supernovae and allow a determination of the distance/redshift
relation to sufficient precision to distinguish between the
effects of an accelerating universe and those of possible
astrophysical phenomena.

\section{Dark Energy and Negative Pressure}

Our current knowledge of the expansion history of the universe can
be summarized as follows: The universe originated at an initial
state that was very close to a density singularity known as the
Big Bang. Soon after that it entered a phase of superluminal
accelerating expansion known as inflation. During inflation
causally connected regions of the universe exited out of the
horizon, the universe approached spatial flatness and the
primordial fluctuations that gave rise to structure were
generated. At the end of inflation the universe was initially
dominated by radiation and later by matter whose attractive
gravitational properties induced a decelerating expansion.

The SnIa data discussed in section 3 (along with other less direct
cosmological observations \cite{Spergel03,Tegmark04}) strongly
suggest that the universe has recently entered a phase of
accelerating expansion at a redshift $z\simeq 0.5$. This
accelerating expansion can not be supported by the attractive
gravitational properties of regular matter. The obvious question
to address is therefore 'What are the properties of the additional
component required to support this acceleration?'. To address this
question we must consider the dynamical equation that determines
the evolution of the scale factor $a(t)$. This equation is the
Friedman equation which is obtained by combining General
Relativity with the cosmological principle of homogeneity and
isotropy of the universe. It may be written as \be \frac{\ddot
a}{a}=-\frac{4\pi G}{3}\sum_i (\rho_i +3p_i)=-\frac{4\pi
G}{3}[\rho_m+(\rho_X + 3p_X)] \label{fe1}\ee where $\rho_i$ and
$p_i$ are the densities and pressures of the contents of the
universe assumed to behave as ideal fluids. The only directly
detected fluids in the universe are matter ($\rho_m,p_m=0$) and
the subdominant radiation ($\rho_r,p_r=\rho_r/3$). Both of these
fluids are unable to cancel the minus sign on the rhs of the
Friedman equation and can therefore only lead to decelerating
expansion. Accelerating expansion in the context of general
relativity can only be obtained by assuming the existence of an
additional component ($\rho_X,p_X=w\rho_X$) termed 'dark energy'
which could potentially change the minus sign of eq. (\ref{fe1})
and thus lead to accelerating expansion. Assuming a positive
energy density for dark energy (required to achieve flatness) it
becomes clear that negative pressure is required for accelerating
expansion. In fact, writing the Friedman eq. (\ref{fe1}) in terms
of the dark energy equation of state parameter $w$ as \be
\frac{\ddot a}{a}=-\frac{4\pi G}{3}[\rho_m+\rho_X (1 + 3w)]
\label{fe1a}\ee it becomes clear that a $w<-\frac{1}{3}$ is
required for accelerating expansion implying repulsive
gravitational properties for dark energy.

The redshift dependence of the dark energy can be easily connected
to the equation of state parameter $w$ by combining the energy
conservation $d(\rho_X a^3)=-p_x d(a^3)$ with the equation of
state $p_X=w\rho_X$ as \be \rho_X\sim a^{-3(1+w)}=(1+z)^{3(1+w)}
\label{rhow}\ee This redshift dependence is related to the
observable expansion history $H(z)$ through the Friedman equation
\be H(z)^2 = \frac{\dot a^2}{a^2}=\frac{8\pi
G}{3}[\rho_{0m}(\frac{a_0}{a})^3+\rho_X(a)]=H_0^2[\Omega_{0m}(1+z)^3+\Omega_X(z)]
\label{fe2}\ee where the density parameter $\Omega\equiv
\frac{\rho}{\rho_{0crit}}$ for matter is constrained by large
scale structure observations to a value (prior) $\Omega_{0m}\simeq
0.3$. Using this prior, the dark energy density parameter
$\Omega_X(z)\equiv \frac{\rho_X(z)}{\rho_{0crit}}$ and the
corresponding equation of state parameter $w$ may be constrained
from the observed $H(z)$.

In addition to $\Omega_X(z)$, the luminosity distance-redshift
relation $d_L(z)$ obtained from SnIa observations can constrain
other cosmological parameters. The only parameter however obtained
directly from $d_L(z)$ (using eq. (\ref{dlh})) is the Hubble
parameter $H(z)$.  Other cosmological parameters can be obtained
from $H(z)$ as follows: \begin{itemize} \item The age of the
universe $t_0$ is obtained as: \be t_0=\int_0^\infty
\frac{dz}{(1+z)H(z)} \ee \item The present Hubble parameter
$H_0=H(z=0)$. \item The deceleration parameter $q(z)\equiv
\frac{{\ddot a}a}{{\dot a}^2}$ \be q(z)=(1+z)\frac{dlnH}{dz}-1 \ee
and its present value $q_0\equiv q(z=0)$. \item The density
parameters for matter and dark energy are related to $H(z)$
through the Friedman equation (\ref{fe2}). \item The equation of
state parameter $w(z)$ obtained as
\cite{Huterer:2000mj,Nesseris:2004wj} \be \label{wz3}
w(z)={{p_X(z)}\over {\rho_X(z)}}={{{2\over 3} (1+z) {{d \ln
H}\over {dz}}-1} \over {1-({{H_0}\over H})^2 \Omega_{0m} (1+z)^3}}
\ee obtained using the Friedman equations (\ref{fe1a}) and
(\ref{fe2}). \end{itemize} The most interesting parameter from the
theoretical point of view (apart from $H(z)$ itself) is the dark
energy equation of state parameter $w(z)$. This parameter probes
directly the gravitational properties of dark energy which are
predicted by theoretical models. The downside of it is that it
requires two differentiations of the observable $d_L(z)$ to be
obtained and is therefore very sensitive to observational errors.

The simplest form of dark energy corresponds to a time independent
energy density obtained when $w=-1$ (see eq. (\ref{rhow})). This
is the well known cosmological constant which was first introduced
by Einstein in 1917 two years after the publication of the General
Relativity (GR) equation \be G_{\mu \nu}=\kappa T_{\mu \nu}
\label{gr1} \ee where $\kappa = 8\pi G/c^2$. At the time the
'standard' cosmological model was a static universe because the
observed stars of the Milky Way were found to have negligible
velocities. The goal of Einstein was to apply GR in cosmology and
obtain a static universe using matter only. It became clear that
the attractive gravitational properties of matter made it
impossible to obtain a static cosmology from (\ref{gr1}). A
repulsive component was required and at the time of major
revolutions in the forms of physical laws it seemed more natural
to obtain it by modifying the gravitational law than by adding new
forms of energy density. The simplest generalization of eq.
(\ref{gr1}) involves the introduction of a term proportional to
the metric $g_{\mu \nu}$. The GR equation becomes \be G_{\mu
\nu}-\Lambda g_{\mu \nu}=\kappa T_{\mu \nu} \label{gr2} \ee where
$\Lambda$ is the cosmological constant. The repulsive nature of
the cosmological constant becomes clear by the metric of a point
mass (Schwarschild-de Sitter metric) which, in the Newtonian limit
leads to a gravitational potential \be
V(r)=-\frac{GM}{r}-\frac{\Lambda r^2}{6} \ee which in addition to
the usual attractive gravitational term has a repulsive term
proportional to the cosmological constant $\Lambda$. This
repulsive gravitational force can lead to a static (but unstable)
universe in a cosmological setup and in the presence of a matter
fluid. A few years after the introduction of the cosmological
constant by Einstein came Hubble's discovery that the universe is
expanding and it became clear that the cosmological constant was
an unnecessary complication of GR. It was then that Einstein
(according to Gamow's autobiography) called the introduction of
the cosmological constant {\it 'the biggest blunder of my life'}.
In a letter to Lemaitre in 1947 Einstein wrote: {\it 'Since I
introduced this term I had always had a bad conscience. I am
unable to believe that such an ugly thing is actually realized in
nature'}. As discussed below, there is better reason than ever
before to believe that the cosmological constant may be non-zero,
and Einstein may not have blundered after all.

If the cosmological constant is moved to the right hand side of
eq. (\ref{gr2}) it may be incorporated in the energy momentum
tensor as an ideal fluid with $\rho_\Lambda= \frac{\Lambda}{8\pi
G}$ and $w=-1$. In the context of field theory such an energy
momentum tensor is obtained by a scalar field $\phi$  with
potential $V(\phi)$ at its vacuum state $\phi_0$ ie $\partial_\mu
\phi=0$ and $T_{\mu \nu}=-V(\phi_0) g_{\mu\nu}$. Even though the
cosmological constant may be physically motivated in the context
of field theory and consistent with cosmological observation there
are two important problems associated with it:
\begin{itemize}\item
{\it Why is it so incredibly small?} Observationally, the
cosmological constant density is 120 orders of magnitude smaller
than the energy density associated with the Planck scale - the
obvious cut off. Furthermore, the standard model of cosmology
posits that very early on the universe experienced a period of
inflation: A brief period of very rapid acceleration, during which
the Hubble constant was about 52 orders of magnitude larger than
the value observed today. How could the cosmological constant have
been so large then, and so small now? This is sometimes called
{\it the cosmological constant problem}. \item {\it The
`coincidence problem':} Why is the energy density of matter nearly
equal to the dark energy density today?
\end{itemize}

Despite the above problems and given that the cosmological
constant is the simplest dark energy model, it is important to
investigate the degree to which it is consistent with the SnIa
data. I will now describe the main steps involved in this
analysis. According to the Friedman equation the predicted Hubble
expansion in a flat universe and in the presence of matter and a
cosmological constant is \be H(z)^2 = \frac{\dot
a^2}{a^2}=\frac{8\pi
G}{3}\rho_{0m}(\frac{a_0}{a})^3+\frac{\Lambda}{3}=H_0^2[\Omega_{0m}(1+z)^3+\Omega_\Lambda]
\label{fecc}\ee where
$\Omega_\Lambda=\frac{\rho_\Lambda}{\rho_{0crit}}$ and \be
\Omega_{0m}+\Omega_\Lambda=1 \label{flt} \ee This is the LCDM
($\Lambda$+Cold Dark Matter) which is currently the minimal
standard model of cosmology. The predicted $H(z)$ has a single
free parameter which we wish to constrain by fitting to the SnIa
luminosity distance-redshift data.

Observations measure the apparent luminosity vs redshift $l(z)$ or
equivalently the apparent magnitude vs redshift $m(z)$ which are
related to the luminosity distance by \be 2.5
log_{10}(\frac{L}{l(z)})=m(z)-M-25=5log_{10}(\frac{d_L(z)_{obs}}{Mpc})
\label{dlobs} \ee From the theory point of view the predicted
observable is the Hubble parameter (\ref{fecc}) which is related
to the theoretically predicted luminosity distance $d_L(z)$ by eq.
(\ref{dlhr}). In this case $d_L(z)$ depends on the single
parameter $\Omega_{0m}$ and takes the form \be
d_L(z;\Omega_{0m})_{th}=c\; (1+z) \int_0^z
\frac{dz'}{H(z';\Omega_{0m})} \label{dlhr1} \ee Constraints on the
parameter $\Omega_{0m}$ are obtained by the maximum likelihood
method \cite{press92} which involves the minimization of the
$\chi^2(\Omega_{0m})$ defined as \be
\chi^2(\Omega_{0m})=\sum_{i=1}^N
\frac{[d_L(z)_{obs}-d_L(z;\Omega_{0m})_{th}]^2}{\sigma_i^2}
\label{chi2} \ee where $N$ is the number of the observed SnIa
luminosity distances and $\sigma_i$ are the corresponding
$1\sigma$ errors which include errors due to flux uncertainties,
internal dispersion of SnIa absolute magnitude and peculiar
velocity dispersion. If flatness is not imposed as a prior through
eq. (\ref{flt}) then $d_L(z)_{th}$ depends on two parameters
($\Omega_{0m}$ and $\Omega_{\Lambda}$) and the relation between
$d_L(z;\Omega_{0m},\Omega_{\Lambda})_{th}$ and
$H(z;\Omega_{0m},\Omega_{\Lambda})$ takes the form \be
d_L(z)_{th}= \frac{c(1+z)}{\sqrt{\Omega_{0m}+\Omega_{\Lambda}-1}}
sin [\sqrt{\Omega_{0m}+\Omega_{\Lambda}-1} \int_0^z dz'
\frac{1}{H(z)}]\nn \label{nfdl} \ee In this case the minimization
of eq. (\ref{chi2}) leads to constraints on both $\Omega_{0m}$ and
$\Omega_{\Lambda}$. This is the only direct and precise
observational probe that can place constraints directly on
$\Omega_\Lambda$. Most other observational probes based on large
scale structure observations place constraints on $\Omega_{0m}$
which are indirectly related to $\Omega_{\Lambda}$ in the context
of a flatness prior.

As discussed in section 2 the acceleration of the universe has
been confirmed using the above maximum likelihood method since
1998 \cite{perl98,riess98}. Even the early datasets of 1998
\cite{perl98,riess98} were able to rule out the flat matter
dominated universe (SCDM: $\Omega_{0m}=1$, $\Omega_{\Lambda}=0$)
at $99\%$ confidence level. The latest datasets are the Gold
dataset ($N=157$ in the redshift range $0<z<1.75$) discussed in
section 2 and the first year SNLS (Supernova Legacy Survey)
dataset which consists of $71$ datapoints in the range $0<z<1$
plus $44$ previously published closeby SnIa. The $68\%$ and $95\%$
$\chi^2$ contours in the ($\Omega_{0m}$ and $\Omega_{\Lambda}$)
parameter space obtained using the maximum likelihood method are
shown in Fig. 11 for the SNLS dataset, a truncated version of the
Gold dataset (TG) with $0<z<1$ and the Full Gold (FG) dataset.
\begin{figure}
\centering
\includegraphics[bb=-40 80 250 680,width=6.0cm,height=10cm,angle=-90]{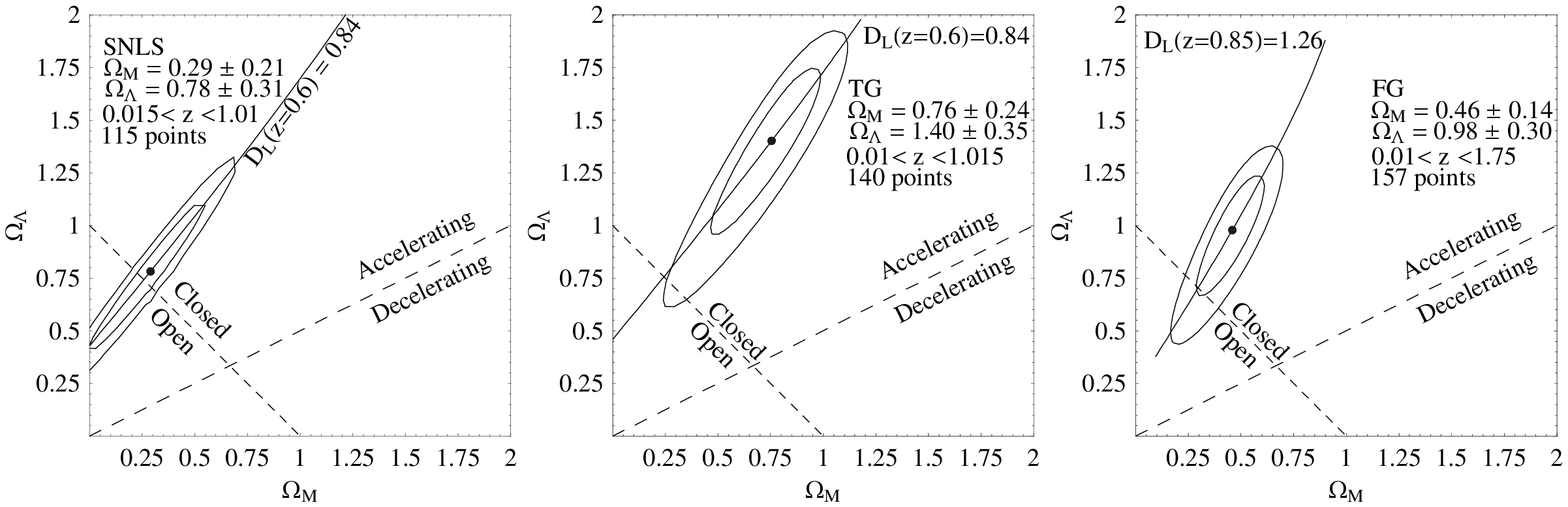}
%\picplace{5cm}{2cm} % Give the correct figure height and width in cm
%
\caption{The $68\%$ and $95\%$ $\chi^2$ contours in the
($\Omega_{0m}$ and $\Omega_{\Lambda}$) parameter space obtained
using the SNLS, TG and FG datasets (from Ref.
\cite{Nesseris:2005ur}).}
\label{fig:11}       % Give a unique label
\end{figure}
The following comments can be made on these plots:
\begin{itemize}
 \item The two
versions of the Gold dataset favor a closed universe instead of a
flat universe ($\Omega_{tot}^{TG}=2.16\pm 0.59$,
$\Omega_{tot}^{FG}=1.44\pm 0.44$). This trend is not realized by
the SNLS dataset which gives $\Omega_{tot}^{SNLS}=1.07\pm 0.52$.
\item The point corresponding to SCDM
$(\Omega_{0m},\Omega_\Lambda)=(1,0)$ is ruled out by all datasets
at a confidence level more than $10\sigma$.  \item If we use a
prior constraint of flatness $\Omega_{0m}+\Omega_\Lambda=1$ thus
restricting on the corresponding dotted line of Fig. 1 and using
the parametrization \be H(z)^2=H_0^2 [\Omega_{0m} (1+z)^2 +
(1-\Omega_{0m})] \ee we find minimizing $\chi^2(\Omega_{0m})$ of
eq (\ref{chi2}) \ba
\Omega_{0m}^{SNLS} &=& 0.26\pm 0.04 \\
\Omega_{0m}^{TG} &=& 0.30\pm 0.05 \\
\Omega_{0m}^{FG} &=& 0.31\pm 0.04 \ea These values of
$\Omega_{0m}$ are consistent with corresponding constraints from
the CMB\cite{Spergel03} and large scale structure
observations\cite{Tegmark04}.
\end{itemize}
Even though LCDM is the simplest dark energy model and is
currently consistent with all cosmological observations
(especially with the SNLS dataset) the question that may still be
address is the following: `Is it possible to get better fits
(lowering $\chi^2$ further) with different $H(z)$ parametrizations
and if yes what are the common features of there better fits?' The
strategy towards addressing this question involves the following
steps:
\begin{itemize} \item Consider a physical model and extract the
predicted recent expansion history $H(z;a_1,a_2,...,a_n)$ as a
function of the model parameters $a_1,a_2,...,a_n$. Alternatively
a model independent parametrization for $H(z;a_1,a_2,...,a_n)$ (or
equivalently $w(z;a_1,a_2,...,a_n)$) may be constructed aiming at
the best possible fit to the data with a small number of
parameters (usually 3 or less). \item Use eq. (\ref{dlhr}) to
obtain the theoretically predicted luminosity distance
$d_L(z;a_1,a_2,...,a_n)_{th}$ as a function of $z$. \item Use the
observed luminosity distances $d_L(z_i)_{obs}$ to construct
$\chi^2$ along the lines of eq. (\ref{chi2}) and minimize it with
respect to the parameters $a_1,a_2,...,a_n$. \item From the
resulting best fit parameter values ${\bar a}_1,{\bar
a}_2,...,{\bar a}_n$ (and their error bars) construct the best fit
$H(z;{\bar a}_1,{\bar a}_2,...,{\bar a}_n)$, $d_L(z;{\bar
a}_1,{\bar a}_2,...,{\bar a}_n)$ and $w(z;{\bar a}_1,{\bar
a}_2,...,{\bar a}_n)$. The quality of fit is measured by the depth
of the minimum of $\chi^2$ ie $\chi_{min}^2({\bar a}_1,{\bar
a}_2,...,{\bar a}_n)$. \end{itemize} Most useful parametrizations
reduce to LCDM of eq. (\ref{fecc}) for specific parameter values
giving a $\chi^2_{LCDM}$ for these parameter values. Let \be
\Delta\chi^2_{LCDM}\equiv \chi_{min}^2({\bar a}_1,{\bar
a}_2,...,{\bar a}_n)-\chi_{LCDM}^2 \ee The value of
$\Delta\chi^2_{LCDM}$ is usually negative since $\chi^2$ is
usually further reduced due to the larger number of parameters
compared to LCDM. For a given number of parameters the value of
$\Delta\chi^2_{LCDM}$ gives a measure of the probability of having
LCDM physically realized in the context of a given parametrization
\cite{Lazkoz:2005sp}. The smaller this probability is, the more
'superior' this parametrization is compared to LCDM. For example
for a two parameter parametrization and $\vert
\Delta\chi^2_{LCDM}\vert
>2.3$ the parameters of LCDM are more than $1\sigma$ away from
the best fit parameter values of the given parametrization. This
statistical test has been quantified in Ref. \cite{Lazkoz:2005sp}
and applied to several $H(z)$ parametrizations.

As an example let us consider the two parameter polynomial
parametrization allowing for dark energy evolution \be
H(z)^2=H_0^2[\Omega_{0m} (1+z)^3 + a_2 (1+z)^2 + a_1 (1+z) +
(1-a_2-a_1-\Omega_{0m})] \label{parpol} \ee in the context of the
Full Gold dataset. Applying the above described $\chi^2$
minimization leads to the best fit parameter values $a_1=1.67 \pm
1.03$ and $a_2=-4.16\pm 2.53$. The corresponding $\vert
\Delta\chi^2_{LCDM}\vert$ is found to be $2.9$ which implies that
the LCDM parameters values ($a_1=a_2=0$) are in the range of
$1\sigma - 2\sigma$ away from the best fit values.

The same analysis can be repeated for various different
parametrizations in an effort to identify the common features of
the best fit parametrizations. For example two other dynamical
dark energy parametrizations used commonly in the literature are
defined in terms of $w(z)$ as
\begin{itemize}
\item Parametrization A: \ba w(z)&=& w_0 +w_1 \; z \label{a1} \\
H^2 (z)&=&H_0^2 [\Omega_{0m} (1+z)^3+\\&+& \nn
(1-\Omega_{0m})(1+z)^{3(1+w_0-w_1)}e^{3w_1 z}]\ea \item
Parametrization B: \ba  w(z)&=&w_0+w_1 \frac{z}{1+z} \label{lda}
\\ H^2 (z)&=&H_0^2 [ \Omega_{0m} (1+z)^3
+ \nn \\ +(1-\Omega_{0m})
(1&+&z)^{3(1+w_0+w_1)}e^{3w_1[1/(1+z)-1]}]\ea
\end{itemize}
where the corresponding forms of $H(z)$ are derived using eq.
(\ref{wz3}). The best fit forms of $w(z)$ obtained from a variety
of these and other parametrizations \cite{Lazkoz:2005sp} in the
context of the Full Gold dataset are shown in Fig. 12.
\begin{figure}
\centering
\includegraphics[bb=40 70 550 790,width=8.0cm,height=12.0cm,angle=-90]{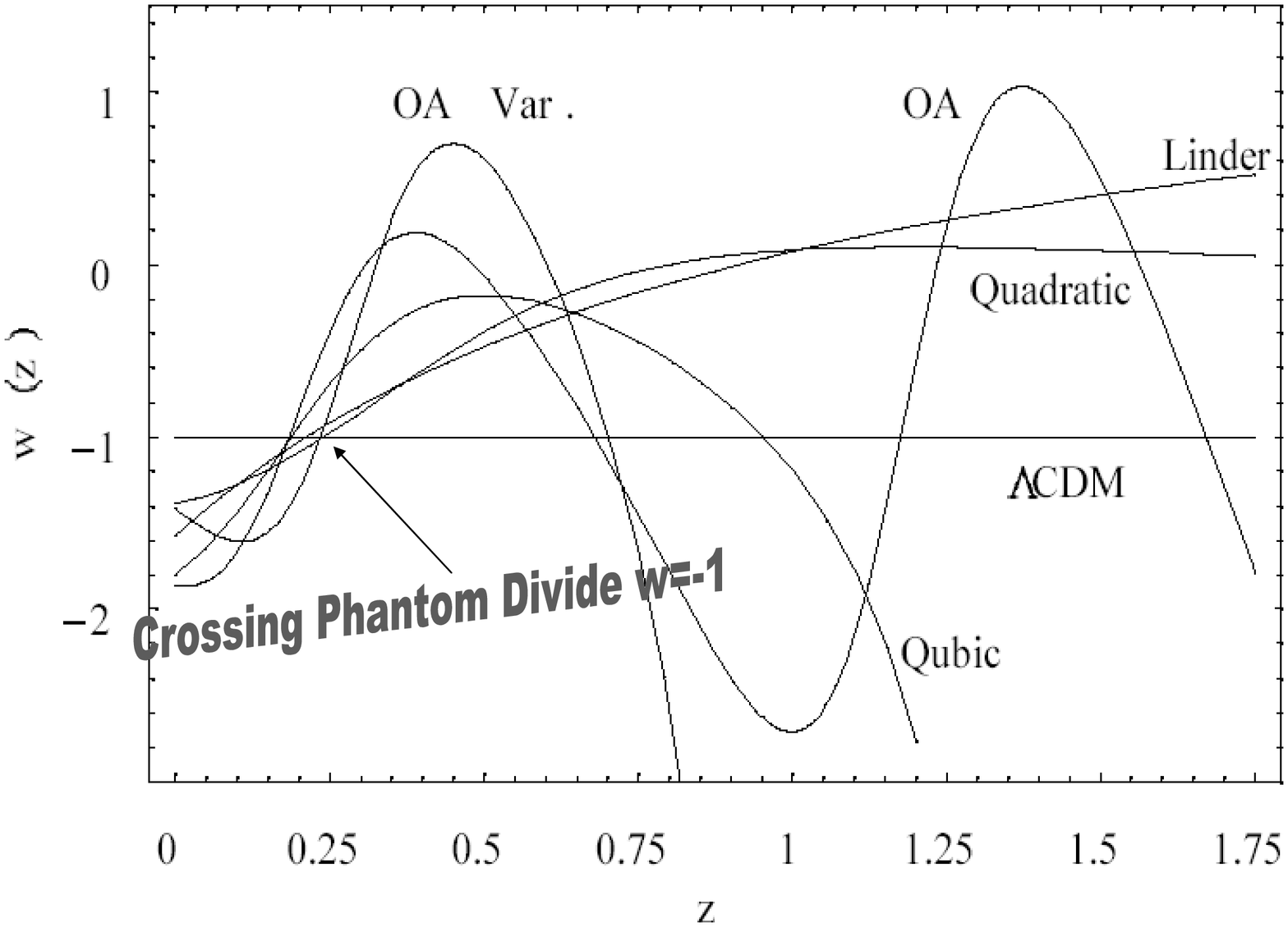}
%\picplace{5cm}{2cm} % Give the correct figure height and width in cm
%
\caption{The best fit forms of $w(z)$ obtained from a variety of
 parametrizations \cite{Lazkoz:2005sp} in the
context of the Full Gold dataset. Notice that they all cross the
line $w=-1$ also known as the Phantom Divide Line (PDL).}
\label{fig:12}       % Give a unique label
\end{figure}
Even though these best fit forms appear very different at
redshifts $z>0.5$ (mainly due to the two derivatives involved in
obtaining $w(z)$ from $d_L(z)$), in the range $0<z<0.5$ they
appear to have an interesting common feature: they all cross the
line $w=-1$ also known as the Phantom Divide Line (PDL). As
discussed in the next section this feature is difficult to
reproduce in most theoretical models based on minimally coupled
scalar fields and therefore if it persisted in other independent
datasets it could be a very useful tool in discriminating among
theoretical models. Unfortunately if the same analysis is made in
the context of the more recent SNLS dataset it seems that this
common feature does not persist.
\begin{figure}
\centering
\includegraphics[bb=30 50 550 570,width=8.0cm,height=12.0cm,angle=-90]{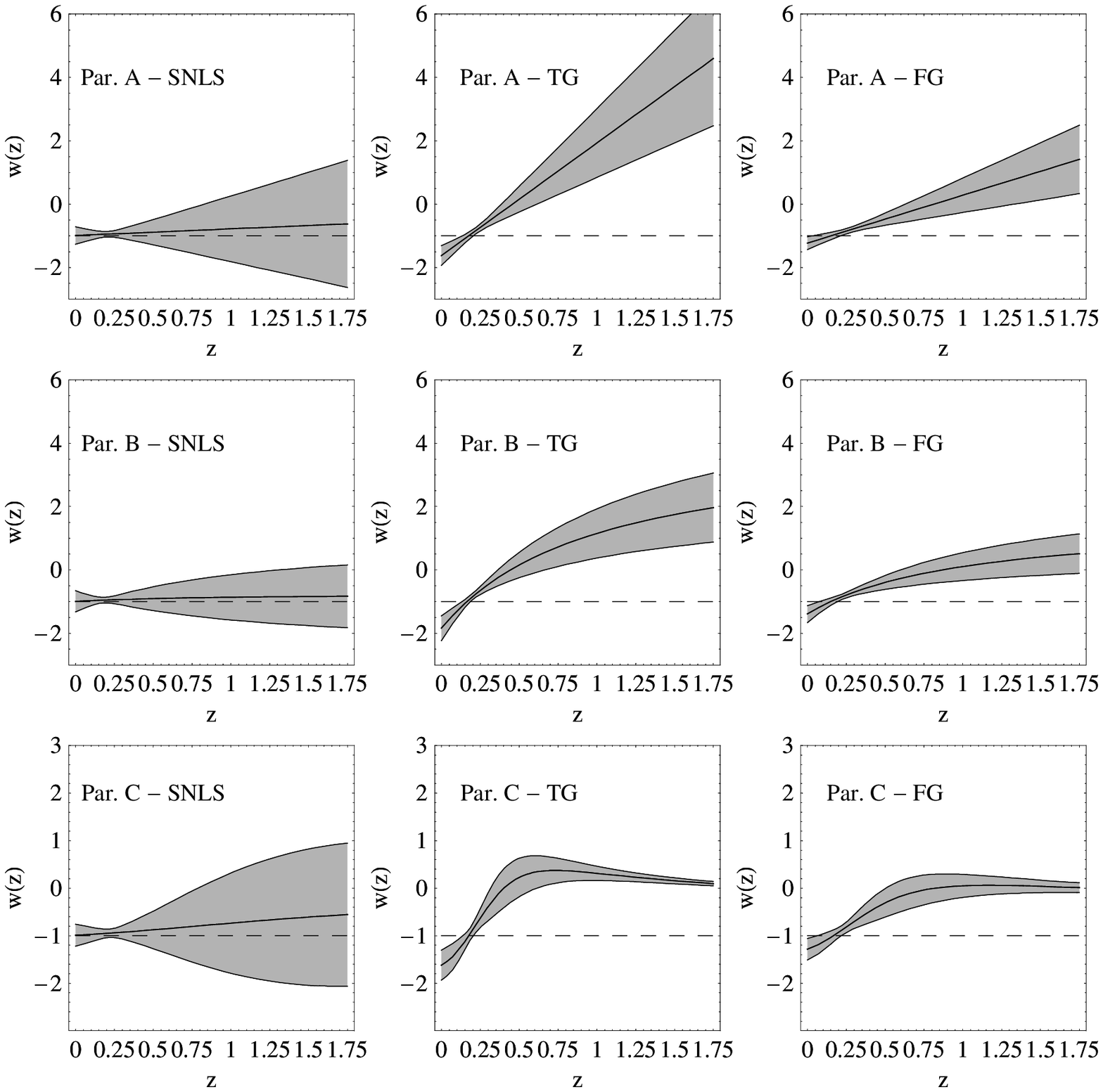}
%\picplace{5cm}{2cm} % Give the correct figure height and width in cm
%
\caption{The best fit $w(z)$ (along with the $1\sigma$ error
(shaded region)) is shown in the context of three different
datasets (in analogy with Fig. 11) for there different
parametrizations (A, B and C) \cite{Nesseris:2005ur}.}
\label{fig:13}       % Give a unique label
\end{figure}
In Fig. 13 the best fit $w(z)$ (along with the $1\sigma$ error
region) is shown in the context of three different datasets (in
analogy with Fig. 11) for the there different parametrizations (A,
B and polynomial of eq. (\ref{parpol}) (called C in Fig. 13)).
Even though the crossing of the PDL is realized at best fit for
both the FG and TG datasets it is not realized at best fit when
the SNLS is used. Thus we must wait until further SnIa datasets
are released before the issue is settled.
\begin{figure}
\centering
\includegraphics[bb=30 70 550 790,width=8.0cm,height=12.0cm,angle=-90]{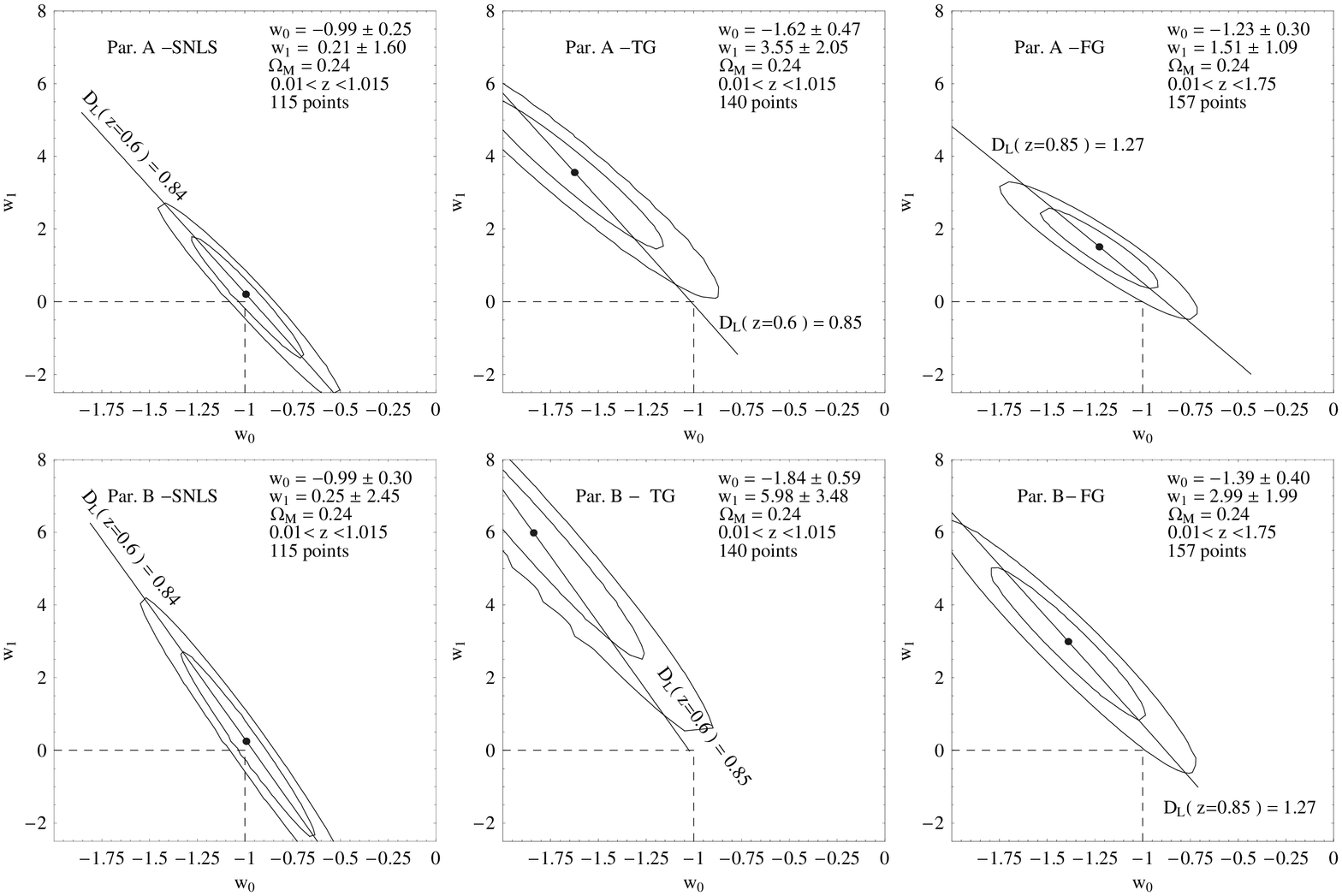}
%\picplace{5cm}{2cm} % Give the correct figure height and width in cm
%
\caption{The $1\sigma$ and $2\sigma$ $\chi^2$ contours
corresponding to parametrizations A and B with a prior of
$\Omega_{0m}=0.24$. Notice that the SNLS dataset provides best fit
parameter values that are almost identical to those corresponding
to LCDM ($w_0=-1$, $w_1=0$).}
\label{fig:14}       % Give a unique label
\end{figure}
In Fig. 14 I show the $1\sigma$ and $2\sigma$ $\chi^2$ contours
corresponding to parametrizations A and B with a prior of
$\Omega_{0m}=0.24$ confirming the fact that the SNLS dataset
provides best fit parameter values that are almost identical to
those corresponding to LCDM ($w_0=-1$, $w_1=0$) despite the
dynamical degrees of freedom incorporated in the parametrizations
A and B. It should be pointed out however that despite the
differences in the best fit parametrizations, the three datasets
(SNLS, TG and FG) are consistent with each other at the $95\%$
confidence range (see e.g. Fig. 14) and they are all consistent
with flat LCDM with $\Omega_{0m}\simeq 0.3$.

\section{Theoretical Models for Dark Energy}

Even though LCDM is the simplest model consistent with current
cosmological data it is plagued with theoretical fine tuning
problems discussed in the previous section (the 'coincidence' and
the 'cosmological constant' problems). In additions dynamical dark
energy parametrizations of $H(z)$ provide in certain cases
significantly better fits to the SnIa data. Therefore the
investigation of physically motivated models that predict a
dynamical evolution of dark energy is an interesting and
challenging problem.

The role of dark energy can be played by any physical field with
positive energy and negative pressure which violates the strong
energy condition $\rho+3p>0$ ($w>-{1\over 3}$). Quintessence
scalar fields\cite{quintess} with small positive kinetic term
($-1<w<-{1\over 3}$) violate the strong energy condition but not
the dominant energy condition $\rho + p>0$. Their energy density
scales down with the cosmic expansion and so does the cosmic
acceleration rate. Phantom fields\cite{phantom} with negative
kinetic term ($w<-1$) violate the strong energy condition, the
dominant energy condition and maybe physically unstable. However,
they are also consistent with current cosmological data and
according to recent
studies\cite{phant-obs2,Nesseris:2004wj,Lazkoz:2005sp} they maybe
favored over their quintessence counterparts.

Homogeneous quintessence or phantom scalar fields are described by
Lagrangians of the form \be  {\cal L}= \pm \frac{1}{2} {\dot
\phi}^2 - V(\phi) \label{lag1} \ee where the upper (lower) sign
corresponds to a quintessence (phantom) field in equation
(\ref{lag1}) and in what follows. The corresponding equation of
state parameter is \be w=\frac{p}{\rho} = \frac{\pm \frac{1}{2}
{\dot \phi}^2 - V(\phi)}{\pm \frac{1}{2} {\dot \phi}^2 + V(\phi)}
\label{eqst1} \ee For quintessence (phantom) models with $V(\phi)
> 0$ ($V(\phi) < 0$) the parameter $w$ remains in the range $-1 <
w < 1 $. For an arbitrary sign of $V(\phi)$ the above restriction
does not apply but it is still impossible for $w$ to cross the PDL
$w=-1$ in a continous manner. The reason is that for $w=-1$ a zero
kinetic term $\pm {\dot \phi}^2 $ is required and the continous
transition from $w<-1$ to $w>-1$ (or vice versa) would require a
change of sign of the kinetic term. The sign of this term however
is fixed in both quintessence and phantom models. This difficulty
in crossing the PDL $w=-1$ could play an important role in
identifying the correct model for dark energy in view of the fact
that data favor $w\simeq -1$ and furthermore parametrizations of
$w(z)$ where the PDL is crossed appear to be favored over the
cosmological constant $w=-1$ according to the Gold dataset as
discussed in the previous section.

It is therefore interesting to consider the available quintessence
and phantom scalar field models and compare the consistency with
data of the predicted forms of $w(z)$ among themselves and with
arbitrary parametrizations of $w(z)$ that cross the PDL. This task
has been recently undertaken by several authors in the context of
testing the predictions of phantom and quintessence scalar field
models\cite{phant-obs2,sn-an}.

As an example we may consider a particular class of scalar field
potentials of the form \be V(\phi) = s \; \phi \label{pot1} \ee
where I have followed Ref. \cite{Garriga:2003nm} and set $\phi =0$
at $V=0$. As discussed in section 2 (see also Ref.
\cite{Garriga:2003nm}) the field may be assumed to be frozen
(${\dot\phi}=0$) at early times due to the large cosmic friction
$H(t)$. It has been argued \cite{Garriga:1999bf} that such a
potential is favored by anthropic principle considerations because
galaxy formation is possible only in regions where $V(\phi)$ is in
a narrow range around $V=0$ and in such a range any potential is
well approximated by a linear function. In addition such a
potential can provide a potential solution to the cosmic
coincidence problem\cite{Avelino:2004vy}.

The cosmological evolution in the context of such a model
\cite{Perivolaropoulos:2004yr} is obtained by solving the coupled
Friedman-Robertson-Walker (FRW) and the scalar field equation \ba
\frac{\ddot a}{a}&=&\mp \frac{1}{3M_p^2}
({\dot \phi}^2 + s \; \phi)-\frac{\Omega_{0m} H_0^2}{2 a^3} \label{fried1} \\
{\ddot \phi} &+& 3 \frac{\dot a}{a}{\dot \phi} - s=0 \label{scal1}
\ea where $M_p = (8\pi G)^{-1/2}$ is the Planck mass and I have
assumed a potential of the form \be V(\phi)=\mp s \; \phi
\label{pot2} \ee where the upper (lower) sign corresponds to
quintessence (phantom) models. The solution of the system
(\ref{fried1})-(\ref{scal1}) for both positive and negative values
of the single parameter of the model $s$, is a straightforward
numerical problem \cite{Perivolaropoulos:2004yr} which leads to
the predicted forms of $H(z;s)$ and $w(z;s)$. These forms may then
be fit to the SnIa datasets for the determination of the best fit
value of the parameter $s$. This task has been undertaken in Ref.
\cite{Perivolaropoulos:2004yr} using the Full Gold dataset. The
best fit value of $s$ was found to be practically
indistinguishable from zero which corresponds to the cosmological
constant for both the quintessence and the phantom cases. The
predicted forms of $w(z)$ for a phantom and a quintessence case
and $s\simeq 2$ is shown in Fig. 15.
\begin{figure}
\centering
\includegraphics[bb=40 70 450 670,width=8.0cm,height=12.0cm,angle=-90]{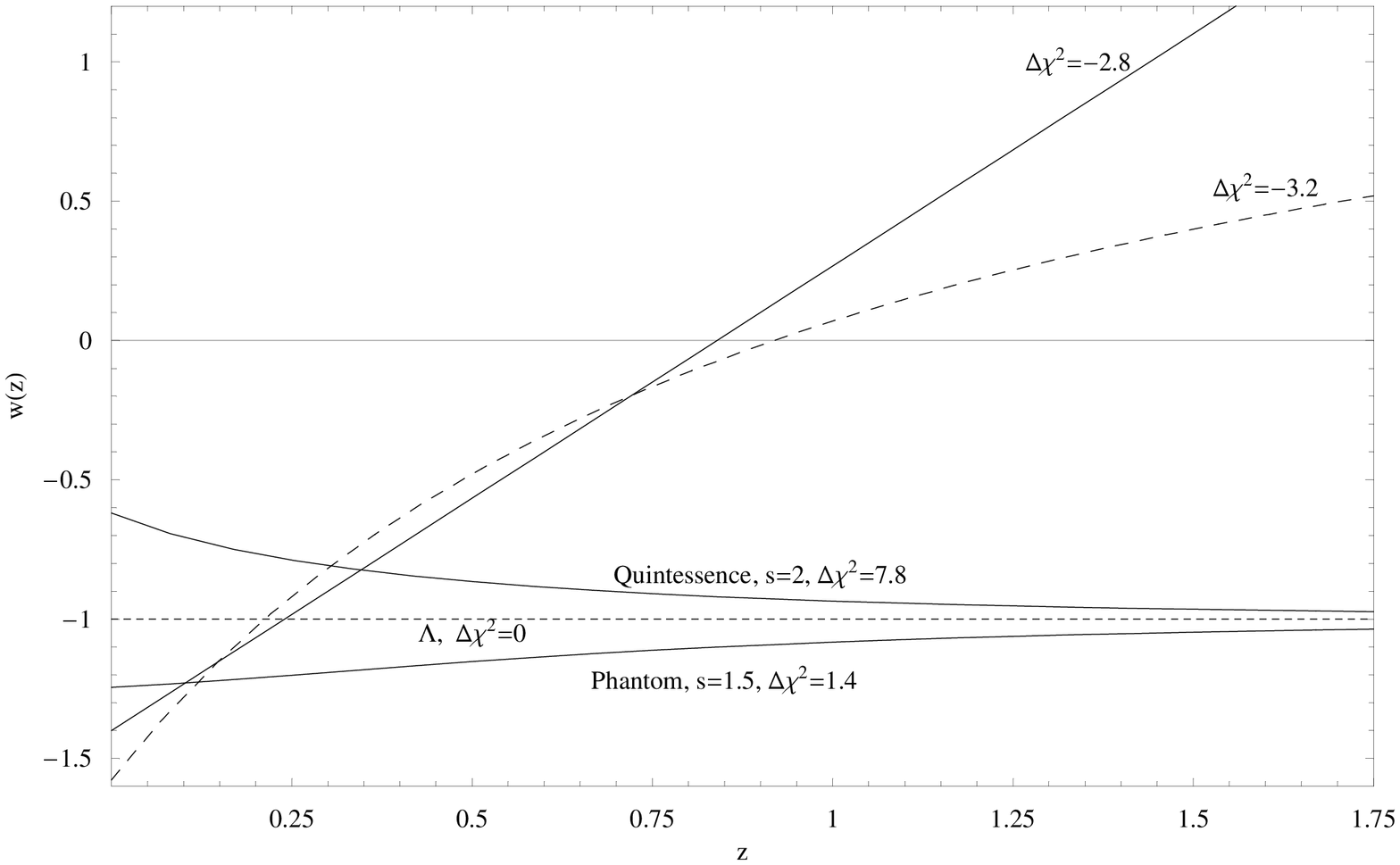}
%\picplace{5cm}{2cm} % Give the correct figure height and width in cm
%
\caption{The predicted forms of $w(z)$ for a phantom and a
quintessence case and $s\simeq 2$ provide worse fits to the Gold
dataset than LCDM and even worse compared to best fit
parametrizations that cross the PDL
\cite{Perivolaropoulos:2004yr}.}
\label{fig:15}       % Give a unique label
\end{figure}
The value of $\Delta\chi^2_{LCDM}$ is positive in both cases which
implies that the fit is worse compared to LCDM. The main reason
for this is that both the quintessence and phantom minimally
coupled scalar field models do not allow for crossing of the PDL
line for any parameter value as discussed above. In contrast, the
best fit $w(z)$ parametrizations A and B of eqs.
(\ref{a1})-(\ref{lda}) which allow for PDL crossing have a
negative $\Delta\chi^2_{LCDM}$ in the context of the Gold dataset
as shown in Fig. 15 and therefore provide better fits than the
field theory models. It should be stressed however that in the
context of the SNLS dataset, parametrizations that allow for
crossing of the PDL do not seem to have a similar advantage as
discussed in the previous section.

The difficulty in crossing the PDL $w=-1$ described above could
play an important role in identifying the correct model for dark
energy in view of the fact that  data favor $w\simeq -1$ and
furthermore parametrizations of $w(z)$ where the PDL is crossed
appear to be favored over the cosmological constant $w=-1$ in the
context of the Gold dataset. Even for generalized k-essence
Lagrangians\cite{Armendariz-Picon:2000ah,Melchiorri:2002ux} of a
minimally coupled scalar field eg \be {\cal L}=\frac{1}{2}f(\phi)
{\dot \phi}^2 - V(\phi) \label{crpdl2} \ee it has been shown
\cite{Vikman:2004dc} to be impossible to obtain crossing of the
PDL. Multiple field Lagrangians (combinations of phantom with
quintessence
fields\cite{Guo:2004fq,Caldwell:2005ai,Hu:2004kh,Stefancic:2005cs})
have been shown to in principle achieve PDL crossing but such
models are complicated and without clear physical motivation (but
see \cite{Nojiri:2005vv} for an interesting physically motivated
model).

The obvious class of theories that could lead to a solution of the
above described problem is the non-minimally coupled scalar
fields. Such theories are realized in a universe where gravity is
described by a scalar-tensor theory and their study is well
motivated for two reasons:
\begin{enumerate}
\item A scalar-tensor theory of gravity is predicted by all
fundamental quantum theories that involve extra dimensions. Such
are all known theories that attempt to unify gravity with the
other interactions (eg supergravity (SUGRA), M-theory etc). \item
Scalar fields emerging from scalar tensor theories (extended
quintessence) can predict an expansion rate $H(z)$ that violates
the inequality \be \frac{d(H(z)^2/H_0^2)}{dz}\geq 3 \Omega_{0m}
(1+z)^2 \label{ineq1} \ee which is equivalent to  crossing the PDL
$w=-1$ (see eg Ref. \cite{Perivolaropoulos:2005yv}).
\end{enumerate}
In fact it has been shown in Ref. \cite{Perivolaropoulos:2005yv}
that in contrast to minimally coupled quintessence, scalar tensor
theories can reproduce the main features of the best fit Hubble
expansion history obtained from the Gold dataset. However, the
precise determination of the scalar tensor theory potentials
requires more accurate SnIa data and additional cosmological
observational input.

\section{The Fate of a Phantom Dominated Universe: Big Rip}
As discussed in section 4 the Gold dataset favors a dynamical dark
energy with present value of the equation of state parameter $w$
in the phantom regime. If this trend is verified by future
datasets and if $w$ remains in the phantom regime in the future
then the fate of the universe acquires novel interesting features.
The energy density of phantom fields increases with time and so
does the predicted expansion acceleration rate ${{\ddot a}\over
a}$. This monotonically increasing acceleration rate of the
expansion may be shown to lead to a novel kind of singularity
which occurs at a finite future time and is characterized by
divergences of the scale factor $a$, the Hubble parameter $H$ its
derivative ${\dot H}$ and the scalar curvature. This singularity
has been called `Big Smash' \cite{McInnes:2001zw} the first time
it was discussed and `Big Rip' \cite{Caldwell:2003vq} in a more
recent study. An immediate consequence of the very rapid expansion
rate as the Big Rip singularity is approached is the dissociation
of bound systems due to the buildup of repulsive negative pressure
in the interior of these systems.

This dissociation of bound systems can be studied by considering
the spacetime in the vicinity of a point mass $M$ placed in an
expanding background in order to study the effects of the cosmic
expansion on bound systems. Such a metric should interpolate
between a static Schwarzschild metric at small distances from $M$
and a time dependent Friedmann spacetime at large distances. In
the Newtonian limit (weak field, low velocities) such an
interpolating metric takes the form\cite{mcvitie}: \be
ds^2=(1-\frac{2GM}{a(t)\rho})\cdot dt^2-a(t)^2\cdot
(d\rho^2+\rho^2\cdot (d\theta^2+sin^2\theta d\varphi^2))
\label{met} \ee

\noindent where $\rho$ is the comoving radial coordinate. Using
\be r=a(t) \cdot \rho \ee the geodesics corresponding to the line
element (\ref{met}) take the form \be -(\ddot{r}-{\ddot{a}\over
a}r)-{GM \over r^2}+r\dot{\varphi}^2=0 \label{geodr} \ee and \be
r^2\dot{\varphi}=L \label{geodf} \ee where $L$ is the constant
angular momentum per unit mass. Therefore the radial equation of
motion for a test particle in the Newtonian limit considered is
\be \ddot{r}={\ddot{a}\over a}r + {L^2 \over r^3}-{GM \over r^2}
\label{radeqm1} \ee The first term on the rhs proportional to the
cosmic acceleration is a time dependent repulsive term which is
increasing with time for $w<-1$. This is easy to see by
considering the Friedman equation (\ref{fe1a}) combined with the
dark energy evolution $\rho_X\sim a^{-3(1+w)}$ where the scale
factor obtained from the Friedman equation is
\begin{equation}
a(t) \,=\,\frac{a(t_m)}{[-w+(1+w)t/t_m]^{-\frac{2}{3(1+w)}}}
\,\,\, for \;\; t >t_m \label{at5}
\end{equation}
and $t_m$ is the transition time from decelerating to accelerating
expansion. For phantom energy ($w<-1$) the scale factor diverges
at a finite time \be t_*=\frac{w}{1+w}t_m >0 \label{brtime} \ee
leading to the Big Rip singularity.
 Clearly, the time dependent repulsive term of eq. (\ref{radeqm1})
diverges at the Big Rip singularity.

A quantitative analysis \cite{Nesseris:2004uj} shows that the
geodesic equation (\ref{radeqm1}) is equivalent to a Newtonian
equation with a time-dependent effective potential  that
determines the dynamics of the bound system which in dimesionless
form is \cite{Nesseris:2004uj} \be
V_{eff}=-\frac{\omega_0^2}{r}+\frac{\omega_0^2}{2r^2}-\frac{1}{2}
\lambda(t)^2 r^2 \label{veff}\ee  where \be
\lambda(t)=\frac{\sqrt{2\vert 1+3w\vert}}{3(-w+(1+w)t)}  \ee with
$w<-1$ and $\omega_0$ is defined as \be
\omega_0^2=\frac{GM}{r_0^3}t_m^2 \ee At $t=1$ the system is
assumed to be in circular orbit with radius given by the minimum
$r_{min}(t)$ of the effective potential of equation (\ref{veff}).
It is easy to show that the minimum of the effective potential
(\ref{veff}) disappears at a time $t_{rip}$ which obeys \be t_* -
t_{rip} = \frac{16 \sqrt{3}}{9} \frac{T \sqrt{2 \vert 1 + 3w
\vert}}{6\pi \vert 1 + w \vert} \label{trip1} \ee The value of the
bound system dissociation time $t_{rip}$ may be verified by
numerically solving the geodesic Newtonian equation of a test
particle with the effective potential (\ref{veff}).
\begin{figure}
\centering
\includegraphics[bb=90 300 500 900,width=8.0cm,height=12.0cm,angle=0]{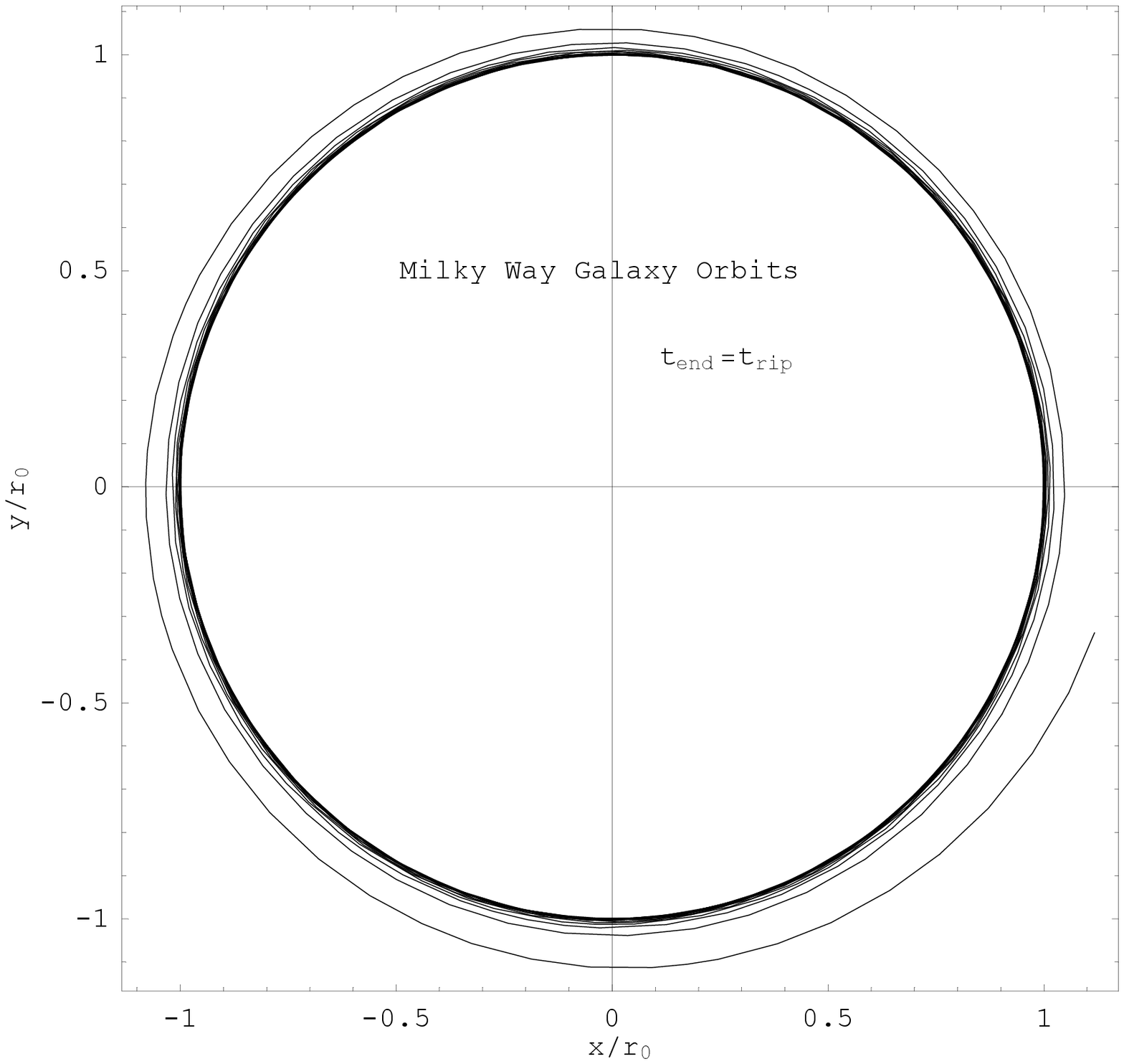}
%\picplace{5cm}{2cm} % Give the correct figure height and width in cm
%
\caption{The numerically obtained evolution of a galactic size two
body system at times close to the predicted dissociation time
$t_{rip}$ \cite{Nesseris:2004uj}.}
\label{fig:16}       % Give a unique label
\end{figure}
The resulting evolution close to the predicted dissociation time
$t_{rip}$ is shown in Fig. 16 for $w=-1.2$ and verifies the
dissociation time predicted by equation (\ref{trip1}).
\begin{table}
\centering \caption{The difference between dissociation times
$t_{rip}$ and the big rip time $t_*$ for three bound systems in
years as predicted by equation (\ref{trip1}). The dissociation
times $t_{rip}$ for the three bound systems in units of $t_m$ are
also shown in column 3. The value $w=-1.2$ was assumed
\cite{Nesseris:2004uj}.}
\label{tab:2}       % Give a unique label
\begin{tabular}{|c|c|c|}\hline
{\bf System }& $t_*-t_{rip}$ (yrs)& $t_{rip}/t_{m}$ \\
\hline {\it Solar System }& $1.88\cdot 10^4 $& $6.00 $\\
\hline {\it Milky Way }& $3.59 \cdot 10^8$& $5.94$  \\
\hline {\it Coma Cluster }& $1.58 \cdot 10^{10}$ & $3.19$ \\
\hline
\end{tabular}
\end{table}
Using the appropriate values for the bound system masses $M$ the
dissociation times of cosmological bound systems may be obtained.
These are shown in Table 2.
\section{Future Prospects-Conclusion}
The question of the physical origin and dynamical evolution
properties of dark energy is the central question currently in
cosmology. Since the most sensitive and direct probes towards the
answer of this question are distance-redshift surveys of SnIa
there has been intense activity during the recent years towards
designing and implementing such projects using ground based and
satellite observatories. Large arrays of CCDs such as MOSAIC
camera at Cerro Tololo Inter-American Obsrevatory, the SUPRIME
camera at Subaru or the MEGA-CAM at the Canada-France-Hawaii
Telescope (CFHT) are some of the best ground based tools for
supernova searches. These devices work well in the reddest bands
(800-900nm) where the ultraviolet and visible light of redshifted
high-z SnIa is detected. Searches from the ground have the
advantages of large telescope apertures (Subaru for example has 10
times the collecting area of the Hubble Space Telescope (HST)) and
large CCD arrays (the CFHT has a 378-milion pixel camera compared
to the Advanced Camera for Surveys on HST which has 16 million
pixels). On the other hand the advantage of space satellite
observatories like the HST include avoiding the bright and
variable night-sky encountered in the near infrared, the potential
for much sharper imaging for point sources like supernovae to
distinguish them from galaxies in which they reside and better
control over the observing conditions which need not factor in
weather and moonlight.

The original two SnIa search teams (the Supernova Cosmology
Project and the High-z Supernova Search Team) have evolved to a
number of ongoing and proposed search projects both satellite and
ground based.
\begin{figure}
\centering
\includegraphics[bb=40 70 550 790,width=8.0cm,height=12.0cm,angle=-90]{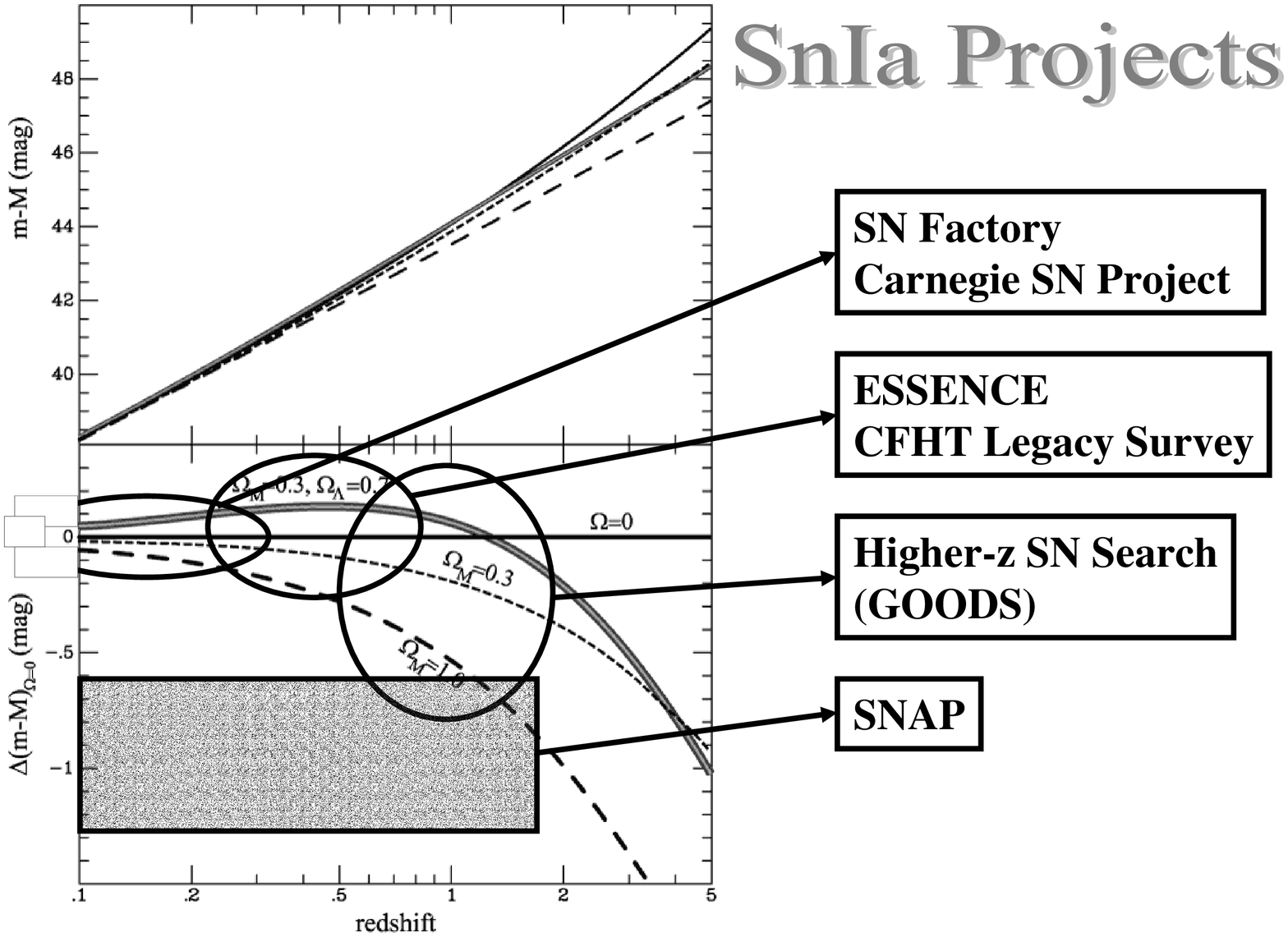}
%\picplace{5cm}{2cm} % Give the correct figure height and width in cm
%
\caption{Ongoing and proposed SnIa search projects with the
corresponding redshift ranges.}
\label{fig:17}       % Give a unique label
\end{figure}
These projects (see Fig. 17) include the following:
\begin{itemize} \item {\bf The Higher-z Supernova Search
Team (HZT)\cite{Strolger:2004kk} and the
GOODS\cite{Giavalisco:2003ig} team of HST:} This has originated
from the High-z Supernova Search Team and has A. Riess of Space
Telescope Sci. Inst. as its team leader. This team is in
collaboration with the GOODS program (Great Observatories Origin
Deep Survey) using the ACS of the HST to detect and analyze high
redshift ($0.5<z<2$) SnIa. Successive GOODS observations are
spaced by 45 days providing 5 epochs of data on two fields: the
Hubble Deep Field (HDF) north and south. Whereas the GOODS team
adds these images to build a superdeep field, the HZT subtracts
the accumulated template image from each incoming frame. Thus the
HZT has already detected more than 42 supernovae in the above
redshift range. \item {\bf Equation of State: SupErNovae trace
Cosmic Expansion (ESSENCE)\cite{Sollerman:2005qj}:} This has also
originated from the High-z Supernova Search Team and has C. Stubbs
of the Univ. of Washington, C. Smith and N. Suntzeff of Cerro
Tololo as its team leaders. This ongoing program aims to find and
measure 200 SnIa's in the redshift range of $0.15<z<0.7$ where the
transition from decelerating to accelerating expansion occurs.
Spectroscopic backup to the program comes from the ground based
Gemini, Magellan, VLT, Keck and MMT Obsevatory. The ESSENCE
project is a five-year endeavor, with the goal of tightly
constraining the time average of the equation-of-state parameter
$w = p/\rho$ of the dark energy. To help minimize systematic
errors, all of their ground-based photometry is obtained with the
same telescope and instrument. In 2003 the highest-redshift subset
of ESSENCE supernovae was selected for detailed study with HST.
\item {\bf The Supenova Legacy Survey
(SNLS)\cite{Palanque-Delabrouille:2005yf}:} The CFHT Legacy Survey
aims at detecting and monitoring about 1000 supernovae in the
redshift range $0<z<1$ with Megaprime at the Canada-France-Hawaii
telescope between 2003 and 2008. High-z spectroscopy of SnIa is
being carried on 8m class telescopes (Gemini, VLT, Keck). Team
representatives are: C. Pritchet (Univ. Victoria), P. Astier
(CNRS/IN2P3), S. Basa (CNRS/INSU) et. al. The SNLS has recently
released the first year dataset \cite{Astier:2005qq}. \item {\bf
Nearby Supernova Factory (SNF)\cite{Wood-Vasey:2004pj}:} The
Nearby Supernova Factory (SNF) is an international collaboration
based at Lawrence Berkeley National Laboratory. Greg Aldering of
Berkeley Lab's Physics Division is the principal investigator of
the SNF. The goal of the SNF is to discover and carefully study
300 to 600 nearby Type Ia supernovae in the redshift range
$0<z<0.3$. \item{\bf Carnegie SN Project
(CSP)\cite{Freedman:2004uz}:} The goal of the project is the
comprehensive study of both Type Ia and II Supernovae in the local
($z < 0.07$) universe. This is a long-term program with the goal
of obtaining exceedingly-well calibrated optical/near-infrared
light curves and optical spectroscopy of over 200 Type~Ia and
Type~II supernovae. The CSP takes advantage of the unique
resources available at the Las Campanas Observatory (LCO). The
team leader is R. Carlberg (Univ. of Toronto).\item {\bf Supernova
Acceleration Probe (SNAP)\cite{Aldering:2002dp}:} This is a
proposed space mission originating from LBNL's Supernova Cosmology
Project that would increase the discovery rate for SnIa's to about
2000 per year. The satellite called SNAP (Supernova / Acceleration
Probe) would be a space based telescope with a one square degree
field of view with 1 billion pixels. The project schedule would
take approximately four years to construct and launch SNAP, and
another three years of mission observations. SNAP has a 2 meter
telescope with a large field of view: 600 times the sky area of
the Hubble Space Telescope's Wide Field Camera. By repeatedly
imaging ~15 square degrees of the sky, SNAP will accurately
measure the energy spectra and brightness over time for over 2,000
Type Ia supernovae, discovering them just after they explode.
\end{itemize} These projects aim at addressing important questions
related to the physical origin and dynamical properties of dark
energy. In particular these questions can be structured as
follows:
\begin{itemize} \item Can the accelerating expansion be attributed
to a dark energy ideal fluid with negative pressure or is it
necessary to implement extensions of GR to understand the origin
of the accelerating expansion? \item Is $w$ evolving with redshift
and crossing the PDL? If the crossing of the PDL by $w(z)$ is
confirmed then it is quite likely that extensions of GR will be
required to explain observations. \item Is the cosmological
constant consistent with data? If it remains consistent with
future more detailed data then the theoretical efforts should be
focused on resolving the coincidence and the cosmological constant
problems which may require anthropic principle arguments.
\end{itemize}
The main points of this brief review may be summarized as follows:
\begin{itemize}
\item {\it Dark energy} with {\it negative pressure} can explain
SnIa cosmological data indicating accelerating expansion of the
universe.\item The existence of a {\it cosmological constant} is
consistent with SnIa data but other {\it evolving} forms of dark
energy {\it crossing the $w=-1$} line may provide better fits to
some of the recent data (Gold dataset). \item New {\it
observational projects} are underway and are expected to lead to
significant progress in the understanding of the properties of
{\it dark energy}.
\end{itemize}
%\begin{theorem}
%Theorem text\footnote{Footnote} goes here.
%\end{theorem}
%
%
% BibTeX users please use
% \bibliographystyle{}
% \bibliography{}
%
% Non-BibTeX users please follow the syntax
% the syntax of "referenc.tex" for your own citations
%\input{referenc}

%%%%%%%%%%%%%%%%%%%%%%%%%%%%%%%%%%%%%%%%%%%%%%%%%%%%%%%%%%%%%%%%%%%%%%  }

%%%%%%%%%%%%%%%%%%%%%%%%%%%%%%%%%%%%%%%%%%%%%%%%%%%%%%%%%%%%%%%%%%%%%%

\printindex
\end{document}